\documentclass[longtitle,preprint,5p,times,twocolumn]{elsarticle}

\usepackage{amssymb,amsmath}

\usepackage{caption}
\usepackage{xspace}
\usepackage{longtable}
\usepackage{pdflscape}
\usepackage{multirow}
\usepackage{ragged2e}
\usepackage{booktabs}
\usepackage{hyperref}
\usepackage{comment}
\usepackage{lipsum}
\usepackage{array}
\usepackage{mfirstuc} 
\usepackage[normalem]{ulem}

\usepackage{adjustbox}
\usepackage{gensymb}

\usepackage{multirow}
\usepackage{multicol}

\newcommand{\Fig}[1]       {Fig.~\ref{#1}}
\newcommand{\Figure}[1]    {Figure~\ref{#1}}

\newcommand{\Equation}[1]  {Equation~\ref{#1}}

\newcommand{\Table}[1]     {Table~\ref{#1}}

\newcommand{\shshlo}{\ensuremath{\sigma_\mathrm{long}^{2}}}
\newcommand{\shshsh}{\ensuremath{\sigma_\mathrm{short}^{2}}}

\usepackage{lineno}

\journal{Nuclear Instruments and Methods A}

\begin{document}

\begin{frontmatter}


\title{Design and Simulated Performance of Calorimetry Systems for the ECCE Detector at the Electron Ion Collider}

\def\theaffn{\arabic{affn}} 
\author[ORNL]{F.~Bock}
\author[ORNL]{N.~Schmidt}
\author[IJCLabOrsay]{P.K.~Wang}
\author[MIT]{N.~Santiesteban}
\author[CUA]{T.~Horn}
\author[BNL]{J.~Huang}
\author[IowaState]{J.~Lajoie}
\author[IJCLabOrsay]{C.~Munoz~Camacho}
\author[MoreheadState]{J.~K.~Adkins}
\author[RIKEN,RBRC]{Y.~Akiba}
% ...
%

% ...
\affiliation[BNL]{organization={Brookhaven National Laboratory},
	 city={Upton},
	 state={NY},
	 country={USA}} 
 
% ...
\affiliation[CUA]{organization={Catholic University of America},
	 city={Washington DC},
	 country={USA}} 
 
% ...
\affiliation[IJCLabOrsay]{organization={Universite Paris-Saclay, CNRS/IN2P3, IJCLab},
	 city={Orsay},
	 country={France}} 
	 
% ...
\affiliation[IowaState]{organization={Iowa State University},
	 city={Iowa City},
	 state={IA},
	 country={USA}} 
 
% ...
\affiliation[MoreheadState]{organization={Morehead State University},
	 city={Morehead},
	 state={KY},
	 }
 
\affiliation[MIT]{organization={Massachusetts Institute of Technology},
	 city={Cambridge},
	 state={MA},
	 country={USA}} 
 
% ...
\affiliation[ORNL]{organization={Oak Ridge National Laboratory},
	 city={Oak Ridge},
	 state={TN},
	 country={USA}} 
 
% ...
\affiliation[RIKEN]{organization={RIKEN Nishina Center},
	 city={Wako},
	 state={Saitama},
	 country={Japan}} 
 
% ...
\affiliation[RBRC]{organization={RIKEN BNL Research Center},
	 city={Upton},
	 state={NY},
	 country={USA}} 
	 
% ...

\begin{abstract}
We describe the design and performance the calorimeter systems used in the ECCE detector design\cite{ecce-paper-det-2022-01} to achieve the overall performance specifications cost-effectively with careful consideration of appropriate technical and schedule risks.
The calorimeter systems consist of three electromagnetic calorimeters, covering the combined pseudorapdity range from -3.7 to 3.8 and two hadronic calorimeters covering a combined range of $-1.1<\eta<3.8$.
Key calorimeter performances which include energy and position resolutions, reconstruction efficiency, and particle identification will be presented.
\end{abstract}

\begin{keyword}
ECCE \sep Electron Ion Collider \sep Tracking \sep Calorimetry 
\end{keyword}

\end{frontmatter}


\setcounter{tocdepth}{1}
\tableofcontents 

\section{Introduction}
\label{sec:introduction}

We report the design and performance of the calorimeter systems for the ECCE detector~\cite{ecce-paper-det-2022-01}.
Homogeneous and sampling calorimeter technologies are employed in the different pseudorapidity regions (backwards, central, and forward) aiming to achieve the overall performance requirements outlined in the EIC Yellow Report~(YR)~\cite{AbdulKhalek:2021gbh} cost effectively and with consideration of technical and schedule risks.
The main physics program of the EIC imposes strong detector performance requirements on the calorimeter systems.
While single inclusive DIS, jets and heavy quark reconstruction require an excellent energy resolution for the electromagnetic and hadronic calorimeters, further requirements for $\pi$ / $e$ separation at the $3\sigma$ level are imposed, for example, by spin asymmetry measurements, TMD evolution, and $XYZ$ spectroscopy.
In order to probe the requested kinematic regions for such processes, a large acceptance in pseudorapidity for the calorimeters is required with special focus on continuous coverage from the backward region to the forward region.
The key performances of the ECCE calorimeter systems are reported and put in context to their impact on physics analyses.
This includes the reconstruction performance, expected energy and position resolution, as well as particle identification via matching to charged particle tracks obtained from the ECCE tracking systems \cite{ecce-paper-det-2022-03}. 

\section {Calorimeter Design}
\label{sec:design}

\begin{figure}[t!]
    \centering
    \includegraphics[width=0.495\textwidth]{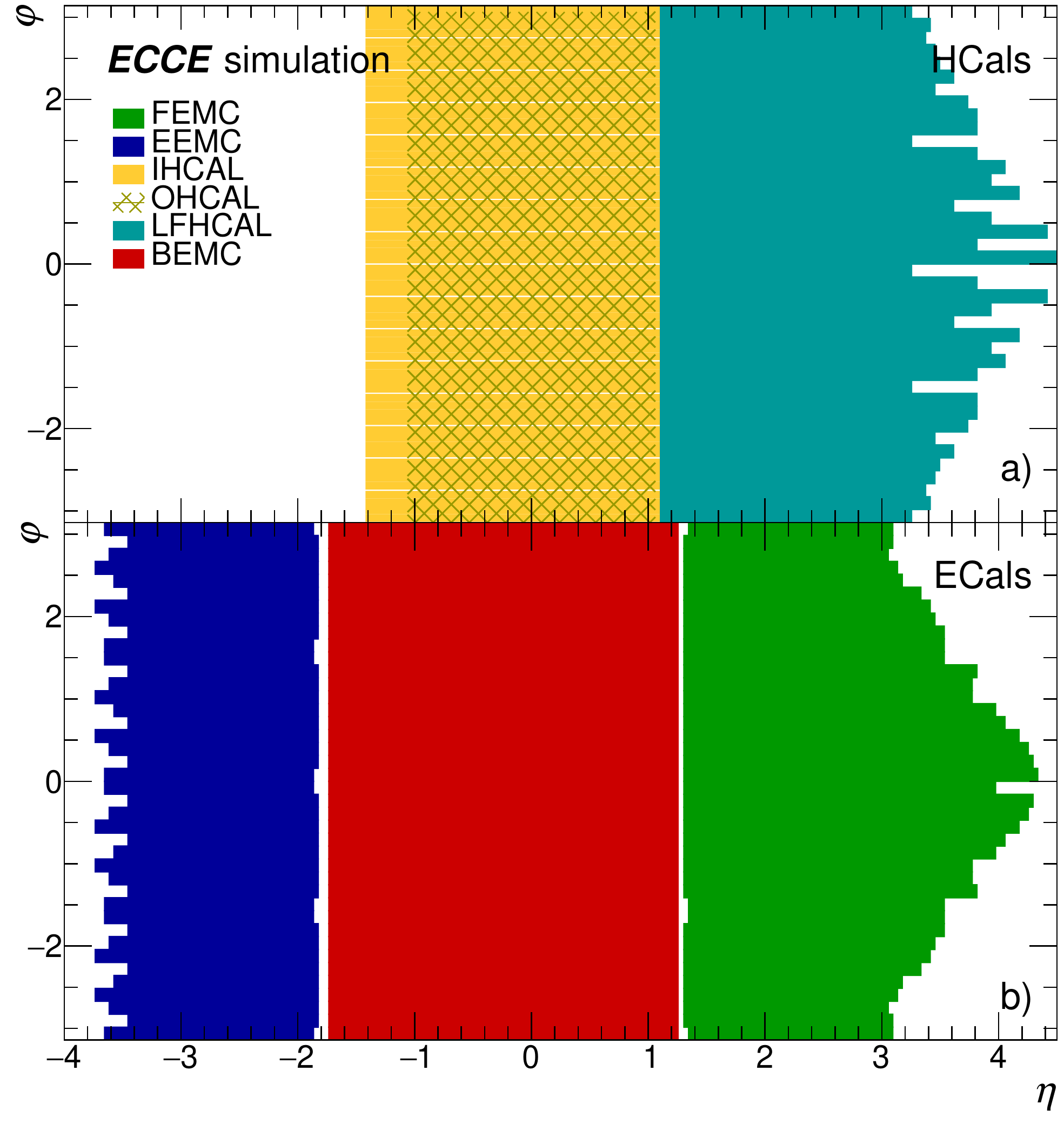}
    \caption{ $\eta-\varphi$ coverage of the ECCE HCals (a) and ECals (b), highlighting the non-uniform tower distributions in azimuth in the forward region due to the crossing angle of the beam pipe.}
    \label{fig:coverage}
\end{figure}
The ECCE calorimeters are designed to address the full range of physics the EIC Whitepaper, the National Academy of Sciences report, and the EIC Yellow Report.
Consequently, particular focus is placed on an excellent electron detection with the broadest possible pseudorapidity ($\eta$) coverage. 
Driven by these concerns, homogeneous electromagnetic calorimeters (ECals) for the electron end cap and the barrel region are selected, while a highly granular shashlik sampling calorimeter is chosen in the hadron going direction. 
The gaps between these calorimeters in $\eta$ are minimized by reducing the support structures for the inner most detectors and even adopting a projective design for the barrel ECal.\\
During the proposal preparations, the ECCE consortium could not identity a physics process which would benefit from a Hadronic Calorimeter (HCal) in the electron end cap. 
Thus, in the presented baseline design for the hadronic calorimeters no HCal is forseen in that direction and instead the two sPHENIX plugdoors will serve as magnet flux return. 
\begin{figure*}[t!]
    \centering
    \includegraphics[width=0.495\textwidth]{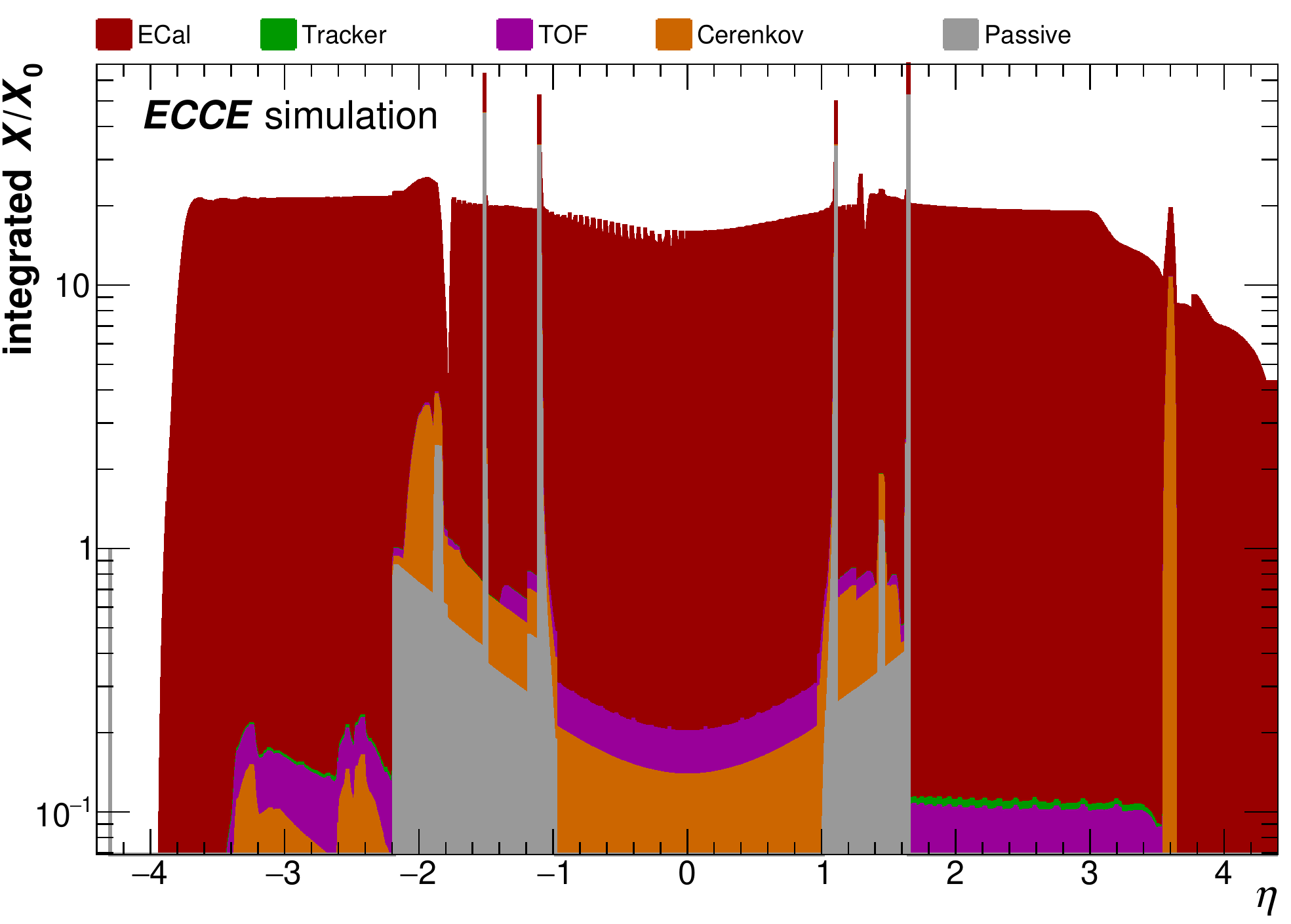}
    \includegraphics[width=0.495\textwidth]{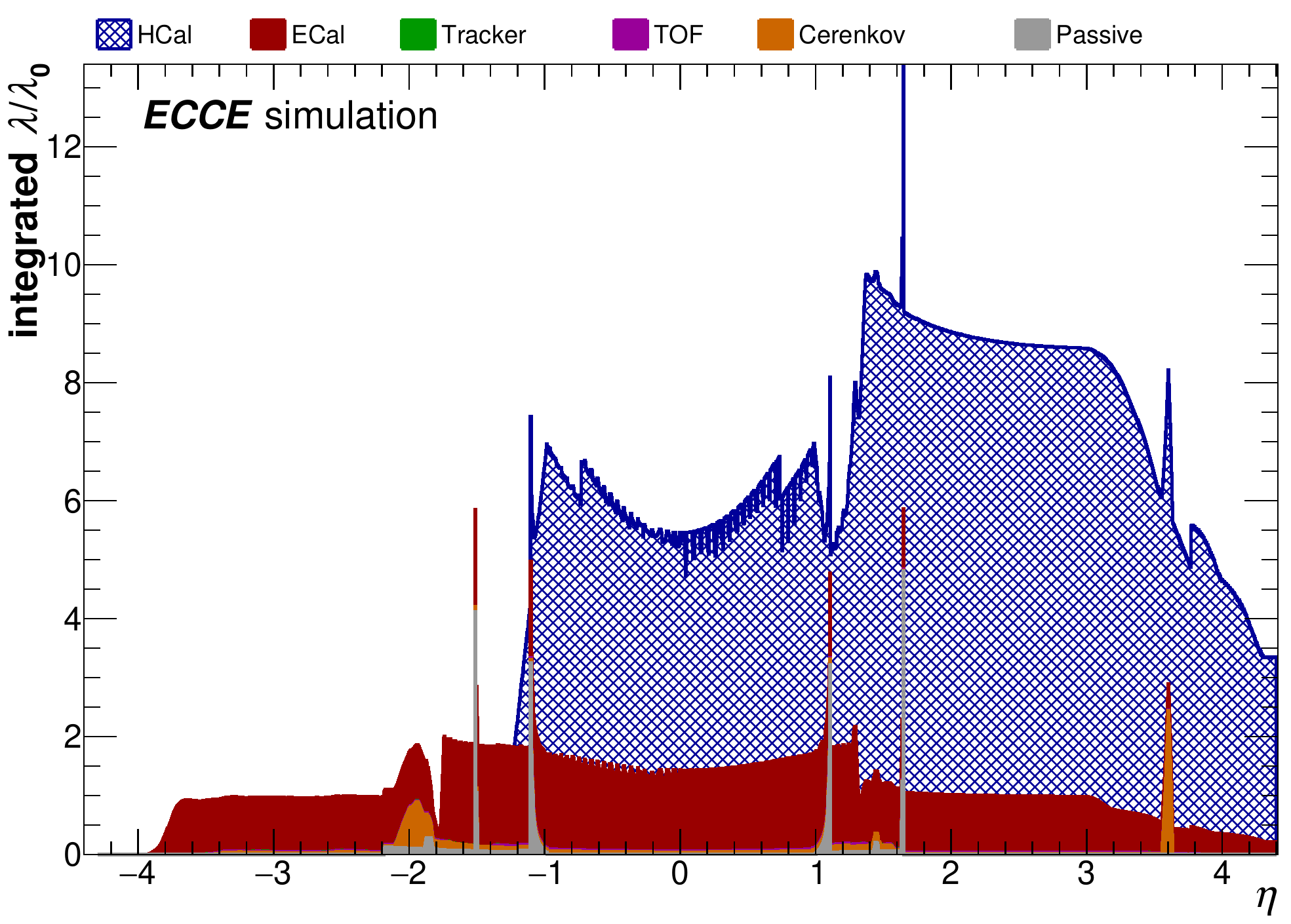}
    \caption{Integrated radiation lengths (left) and nuclear interaction lengths (right) in the full ECCE detector configuration as a function of pseudorapdity. Contributions of individual detector systems are summed up according to six material categories.}
    \label{fig:matbudsums}
\end{figure*}
For the barrel we propose to repurpose the existing outer HCal from the sPHENIX collaboration, which is currently under construction at BNL \cite{sPHENIX:tdr}. 
This HCal surrounding the BaBar magnet will be complemented by an instrumented steel support frame that holds the barrel ECal. 
Despite its limited depth, this inner HCal will be able to serve as calibration point before the magnet.
In the hadron going direction we propose to construct a new longitudinally separated HCal in order to capture the rather collimated hadrons going in this direction with the best possible energy resolution. 
The acceptance of the envisioned detectors in $\eta$ and azimuth ($\varphi$) according to the ECCE GEANT4 implementations for all HCals (top) and ECals (bottom), can be found in \Figure{fig:coverage}.
The figure also shows that the calorimeters cover the full azimuth ($0<\varphi<2\pi$) in most of the pseudorapdity regions.
In the forward region, the 25 mrad crossing angle of the beam pipe results in a $\varphi$-asymmetric setup in particular at high pseudorapidities.

The performance of the above described calorimeters strongly depends on the material budget of the inner detectors, as early material interactions can deteriorate the reconstruction performance. 
A special focus here is put on the ECals where excess material of the inner detectors could quickly add up to several percent of a radiation length ($X/X_0$).
Thus, the material of all inner detector systems and support frames in ECCE has been greatly minimized by design, resulting in a material budget of only $0.2-1X/X_0$ in the barrel and approximately $0.15X/X_0$ in the forward and backward direction with slight modulations depending on $\eta$.
The corresponding $\eta$-distribution of $X/X_0$ including the ECals is shown in \Fig{fig:matbudsums}~(left), while the respective distribution of the nuclear interaction length ($\lambda/\lambda_0$) including the HCals can be found in \Fig{fig:matbudsums}~(right).
As can be seen, the bulk of material in front of the ECals stems from the Cherenkov (mRICH, dRICH, DIRC) detector systems and from the TOF systems. 
The $\eta$ regions between the barrel and forward/backward calorimeters shows several significant passive support structures in the distribution, whose material we aim to reduce considerably.
For the HCals, the bulk of upstream material is given by the ECals as well as by the passive magnet material in the barrel.
The final number of nuclear interaction lengths and radiation lengths of the different calorimeters that are described in this article can also be obtained from \Fig{fig:matbudsums}, which is based on a GEANT4 material scan of the full ECCE detector as implemented in the Fun4All framework \cite{ecce-note-comp-2021-01-nim}.

\subsection{Electron-End-Cap Electromagnetic Calorimeter: EEMC}
The electron-end-cap calorimeter will to cover a dynamic energy range of 0.1--18 GeV for electromagnetic showers of the scattered electron based on e+p Pythia simulations at 18x275 GeV$^2$. 
The choice of technology and detector dimensions are therefore optimized to provide the optimal performance for this expected energy range.

The EEMC is a high-resolution ECal designed for precision measurements of the energy of scattered electrons and final-state photons in the electron-going region. 
The requirements for energy resolution in the backward region is driven by inclusive DIS where precise determination of the scattered electron properties is critical to constrain the event kinematics. 
The EEMC is designed to address the requirements outlined in the EIC Yellow Report.
Its baseline design is based on an array of approximately 3000 lead tungsten crystals (PbWO$_{4}$) $2\times 2\times 20$~cm$^3$ in size, which correspond to approximately 20~$X/X_0$ longitudinally and a transverse size equal to the PbWO$_{4}$ Moli\`ere radius. 
The PbWO$_{4}$ crystal light yield is in the range of 15 to 25 photo-electrons per MeV, providing an excellent energy resolution of $\sigma_E/E\approx2\%/\sqrt{E} \oplus 1\%$ \cite{Inaba:1994jd,Prokoshkin:1995rd} within a very compact design.\\
The EEEMCAL Consortium is leading the efforts to further develop the EEMC design concept and has summarized their intentions in an Expression Of Interest in 2021.
They have begun to organize activities into mechanical design, scintillator, readout, and software/simulation among the collaborating institutions. 
Pre-design activities of the mechanical support structure commenced in 2021 and a document on mechanical design and integration has been prepared \cite{eeemcal:design}. 
The concept is based existing detectors that the team has constructed, and in particular the Neutral Particle Spectrometer at Jefferson Lab. \\
\begin{figure}[!t]
    \centering
    \includegraphics[width=\linewidth]{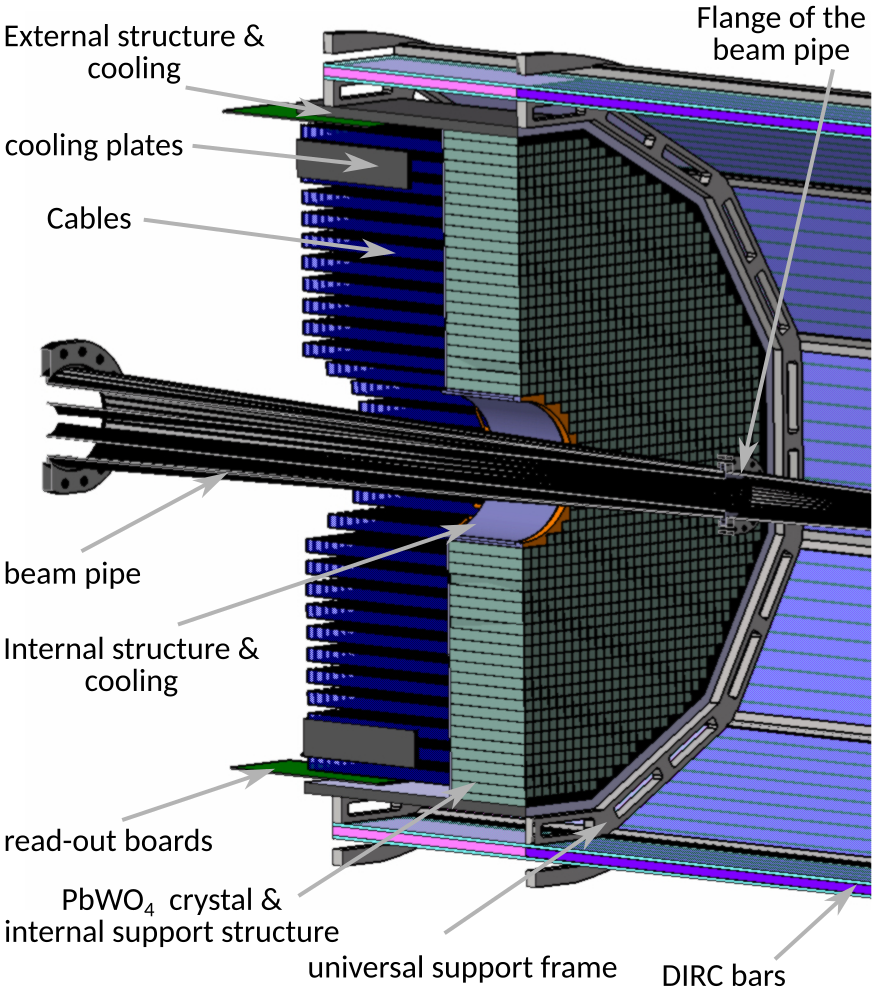}
    \caption{The Electron-End-Cap Calorimeter conceptual design and integration with the beampipe and surrounding detectors as prepared by the EEEMCAL Consortium \cite{eeemcal:design}. 
    The EEMC consists only of PbWO$_{4}$ crystals and uses the displayed design concept. 
    }
    \label{fig:eemcal_config}
    \label{fig:eemcal_clearance}
\end{figure}
\Figure{fig:eemcal_config} shows an overview of the different components of the EEMC as prepared by the EEEMCAL Consortium \cite{eeemcal:design}. 
It has four main parts: the detector (PbWO$_{4}$ crystals), the mechanical structure (internal and external), cooling, and electronics (SiPM and cables). 
With crystal dimensions of $2\times2\times20$~cm$^3$, a density of 8.28~g/cm$^3$, and a mass of 0.6624~kg per crystal the total weight of the EEMC is slightly more than two metric tons. 
The crystals are aligned and separated using carbon plates of thickness 0.5~mm. 
The configuration for the first ring of PbWO$_{4}$ crystals depends on the final design of the beam pipe. 
Its minimum diameter will be on the order of 22.5~cm with an additional clearance gap. 
An additional support and cooling structure with a maximum thickness of 5~mm will be needed to support the crystals directly above the beam pipe.
For which detailed calculations will have to be carried out once the final beam pipe design is available.
The EEMC is located inside the universal support frame, which also houses the Detection of Internally Reflected Cherenkov light (DIRC) detector \cite{ecce-paper-det-2022-01}, and covers the pseudorapdity region of $-3.4<\eta<-1.5$. 
The main constraints for its acceptance are imposed by the surrounding detector systems and passive materials, as seen in \Fig{fig:eemcal_config}.
The integration of the EEMC into the frame is only possible if the beam pipe is removed, which implies that the flange must be disconnected. 
To improve the inner diameter of the EEMC and to improve the acceptance up to $-3.7<\eta<-1.5$, an inner calorimeter is being considered. 
This option also requires the modification of the overall structure of the EEMC to ensure no significant gaps in scattered electron detection between the electron-end-cap and barrel.
Overall, the inner diameter of the EEMC will depend on the design of the beam pipe, and in particular the angle between the electron and the hadron tube.\\
Currently, the EEMC readout is based on silicon photomultipliers (SiPMs) of pixel sizes of 10$\mu$m or 15$\mu$m and a photosensitive area of $3\times3$~mm$^2$. 
There are two configuration options: 4 SiPM per crystal or 16 SiPM per crystal. 
Since a mechanical structure is required for mounting the PCBs, its width in turn will determine the positioning of the SiPMs.
Assuming a machined grid with a width of about 5~mm the PCBs can be mounted with small screws. \\
PbWO$_{4}$ crystals are sensitive to temperature changes with a variation of 2\%/$^\circ$C in light output. 
Thus, the specification is to keep the crystal temperature stable within $\pm$0.1 $^\circ$C.
To ensure this stability the additional heat generated by the electronics needs to be removed and the following cooling structures are being considered.
As internal cooling structure several machined copper blocks with internal coolant circulation will be used around the beam pipe.
To reduce the spatial extend support structures the EEEMC consortium is moreover planning to use cooling plates in between the readout cables which are linked to the support structure surrounding the EEMC with tubes.
This system is composed of 12 plates with a 5-8~mm spacing in which water can be circulated. 
The cooling near the crystals will likely not be enough to meet specification.
These challenges could be overcome by outside cooling with standard cooling blocks with airflow in front of the electronics or additional cooling added at the back of the assembly. 
The main constraint is the space available in the electron end-cap. \\
The mechanical integration of the EEMC presently envisions that the detector is assembled in its own support structure, mounted on a platform, and then inserted into the universal support frame. 
The detailed steps and main points of the assembly are described in Ref.~\cite{eeemcal:design}. 
The mechanical integration starts when the assembly is complete. 
The platform is adjusted on rails with an additional support to link the support to the detector. 
The platform is removed once the EEMC is mounted in the universal support frame. 
Clearances of at least 5~mm on all sides between the EEMC and the universal support frame are required to perform maintenance without lifting the detector. 
\begin{figure*}[ht!]
    \centering
    \includegraphics[height=0.22\textheight]{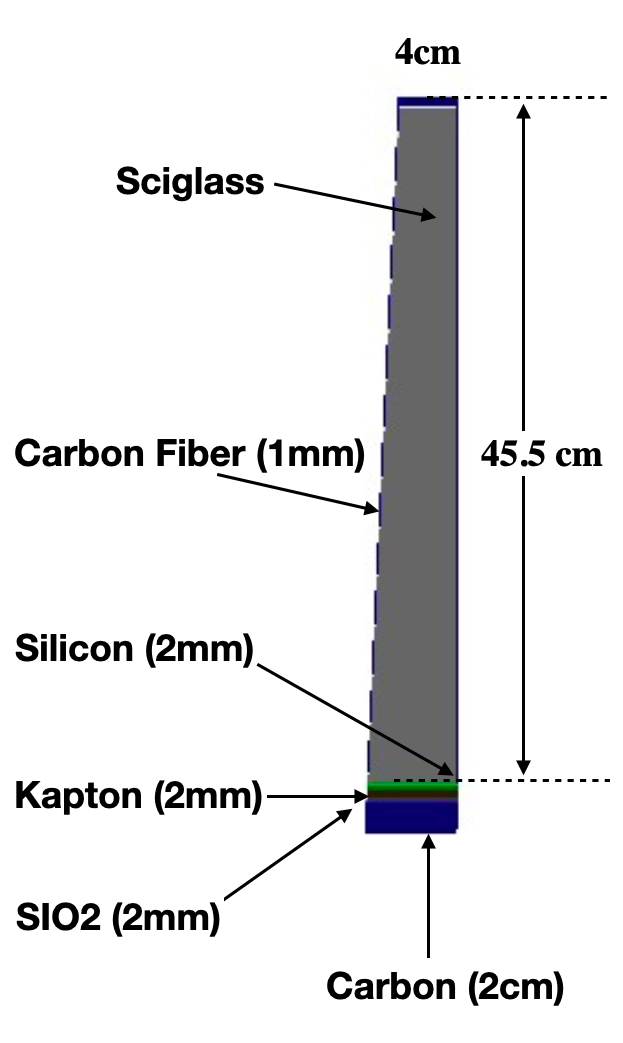}
    \includegraphics[height=0.22\textheight]{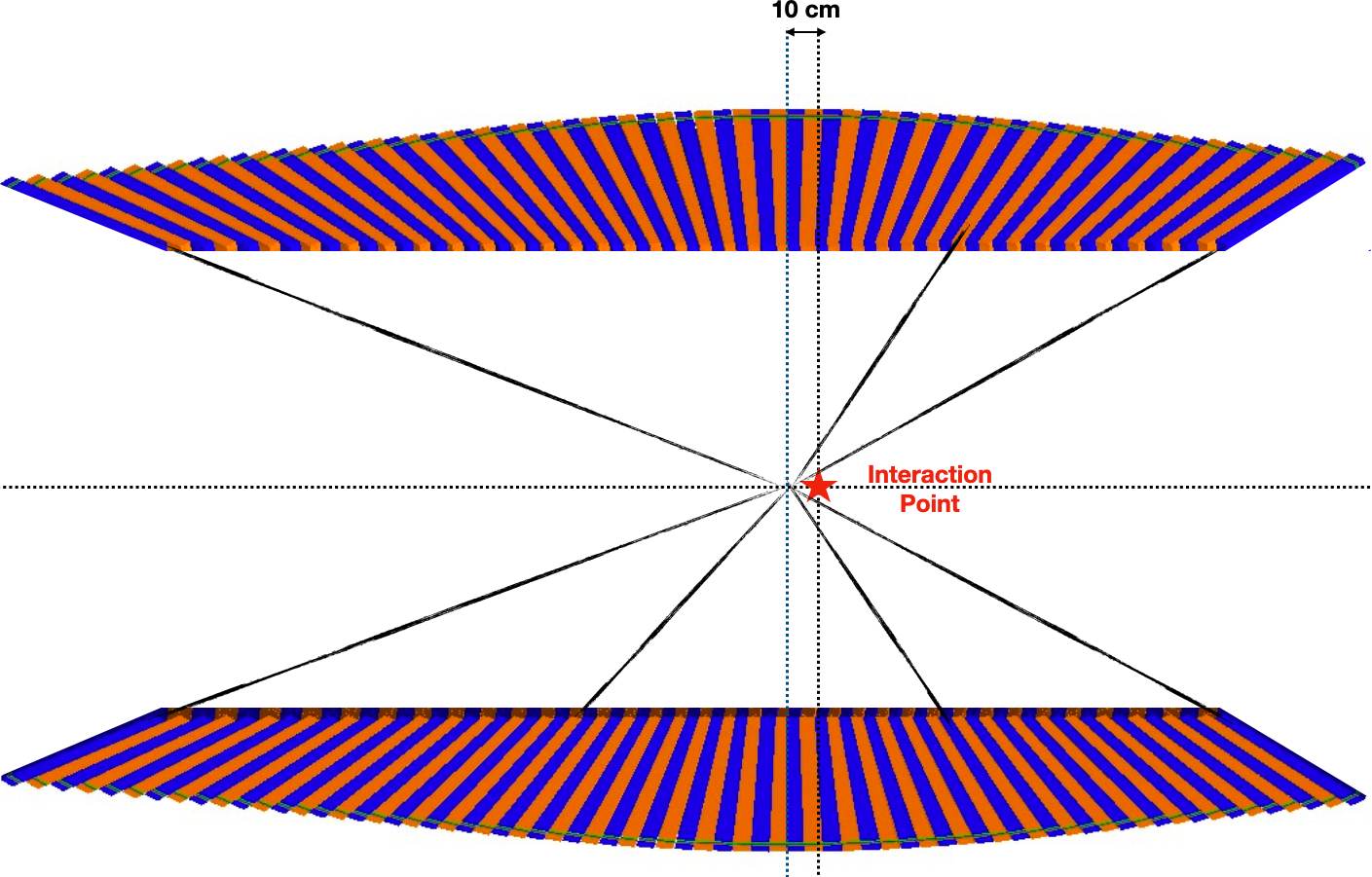}
    \includegraphics[height=0.22\textheight]{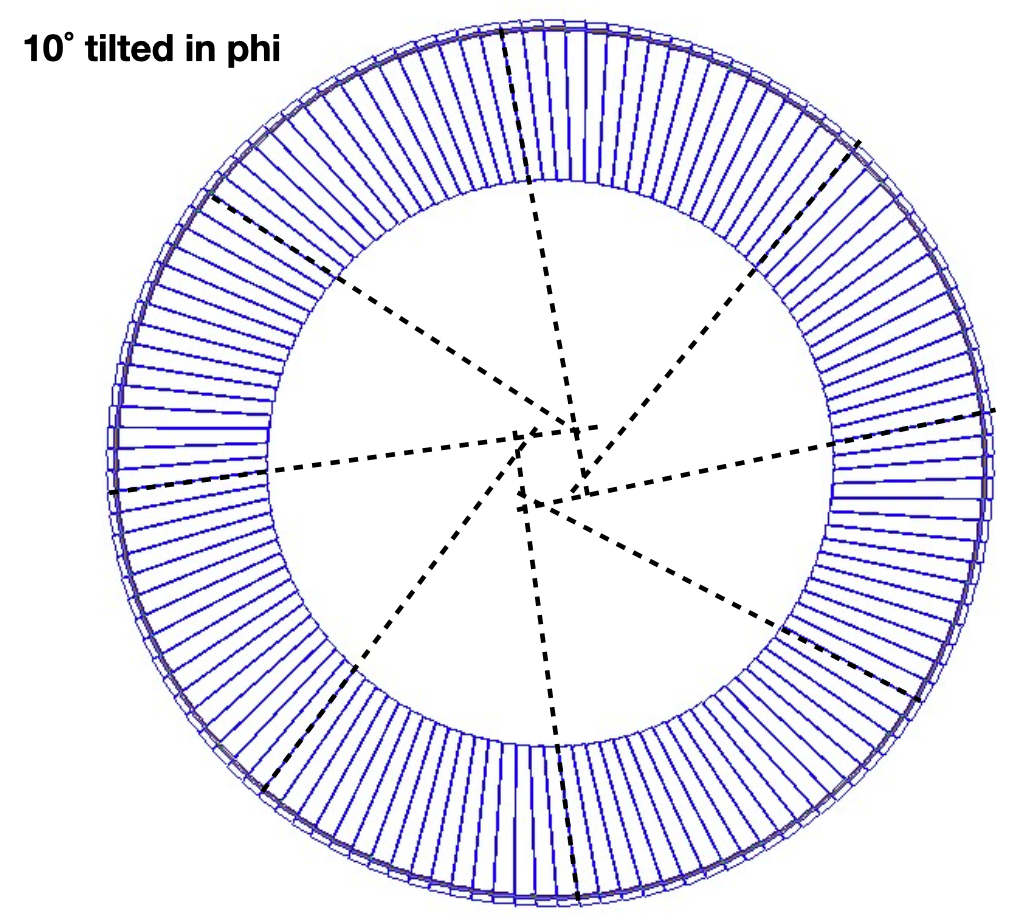}
    \caption{Left: Single BEMC tower as implemented in the GEANT4 simulations. 
    Middle: BEMC projective tower layout in $\eta$. 
    The towers are centered at $z = -10$~cm. 
    Right: BEMC layout layout as a function of $\varphi$ for one $\eta$ slice.
    The towers are tilted in $\varphi$ by $10^\circ$ to avoid channeling in the gaps between adjacent towers.
    }
    \label{fig:becal_eta}
    \label{fig:becal_phi}
    \label{fig:becal_tower}
\end{figure*}

\subsection{Barrel}
Based on Pythia simulations of e+p collisions at 18x275 GeV$^2$, the expected energy range of particles at mid-rapidity is 0.1--50 GeV in high $Q^2$ events. 
The ECCE detector therefore requires calorimeters that can cover these expected energies for electromagnetic and hadronic shower reconstruction.
\paragraph{Barrel Electromagnetic Calorimeter: BEMC} \mbox{ }\\
The barrel electromagnetic calorimeter is designed to cover the central region of the detector ($-1.72 < \eta <1.31$). 
Its total length along the $z$-axis is 584~cm and the detector is fully contained within solenoid magnet, but positioned at a larger radial position than the DIRC detector.
The absolute radial position of the calorimeter is $85<R< 135$~cm from the beampipe, where the inner radius is fixed for all towers but the outer radius varies depending of the position in eta due to the required projective design. \\
The calorimeter is composed of 8960 towers made out of scintillating glass (SciGlass), which are organized in 128 towers per $\varphi $ slice and 70 blocks in the $\eta$ direction. 
The middle and left panels of \Fig{fig:becal_eta} show $x$ and $z$ slices of the BEMC geometry as it is modeled in GEANT4. 
The colors show the different $\eta $ towers, and the variation in the outer radius can be observed. 
The interaction point is marked by a red star, from which the left and right directions in the figure correspond to the electron and hadron-going sides, respectively. 
The detector is designed asymmetric towards the electron and hadron sides, resulting in a length of 389 and 195~cm on the respective sides.
The BEMC is designed with offset projectivity in $\eta $ and $\varphi$ in order to avoid channeling of particles produced in the collision within the passive material between the towers.
Thus the front face of the towers is tilted such that it is facing the interaction point shifted by $\Delta z = -10$~cm.
This requires that the tower tilt angle depends on its location within the calorimeter.
Additionally, the towers have a stronger inclination at higher absolute pseudorapidities, leading to an asymmetric tapered shape of the glass blocks, which increases with $|\eta|$. 
All the towers are tilted by $\Delta\varphi=10 ^{\circ}$ to avoid any gaps in $\varphi$ and further tunneling of particles through inactive detector material. 
A summary of all BEMC detector parameters is given in Table~\ref{tab:becalpar}.\\
\begin{table}[!t]
  \centering
  \begin{tabular}{lr}
    \toprule
    Parameter & Value \\
    \midrule
    Inner radius (envelope)   & 85 cm \\
    Outer radius (envelope)    & max. 135 cm ($\eta-\text{dependent})$\\
    Length (envelope)           & 584 cm ($-389<z<195$ cm) \\
    Pseudorapidity coverage & $-1.72<\eta<1.31$\\
    Active material                &  SciGlass \\ 
    
    \# towers in azimuth        & 128 \\
    \# towers in pseudorapidity & 70 \\
    Tower dimensions \\
    \quad inner face: & $4\times4$ cm\\
    \quad length: & $45.5$ cm\\
    \quad outer face ($\eta=0$): & $5\times5$ cm\\
    \quad outer face ($|\eta|>1.1$): & $6.6\times6.6$ cm\\
    $\eta$ projectivity point & $z=-10$ cm \\
    $\varphi$ projectivity tilt & $10^\circ$ \\
    Sampling fraction & 0.97\\
    Tower depth & $X/X_0\approx16.0$ \\
    Moli\`ere radius & $R_\mathrm{M}=3.58$ cm\\
    \bottomrule
  \end{tabular}
  \caption{Design parameters for the barrel electromagnetic calorimeter BEMC.}
  \label{tab:becalpar}
\end{table}
The layout for a single tower around $\eta = 0$ is shown in \Fig{fig:becal_tower}~(left).
All towers currently have an inner size of $4 \times 4$~cm$^2$ and the same length of 45.5~cm, which corresponds to approximately 16 $X/X_0$.
However, their outer face dimension varies from $5^2$ to $6.6^2$~cm$^2$ depending on their position in $|\eta|$.
In addition, the considered SciGlass towers have a Moli\`ere radius of 3.58~cm, which is approximately double the transverse tower size.
Each tower is composed of a SciGlass core, surrounded by a 1~mm carbon fiber enclosure. 
The electronics are currently modeled by Kapton, SiO$_{2}$ and carbon fiber layers in the outer part of the blocks. 
The SciGlass block length is optimized to contain at least $95\%$ of the energy of a 10 GeV electron, whilst still fitting into the BABAR 1.5T magnet with at most an inner radius of 80~cm and at least 8~cm space for the electronics and support structure.
The electron energy mentioned above corresponds to the average scattered electron energy in the BEMC acceptance. 
Constraining the BEMC to not stretch further into the detector allows for more space for other PID and tracking detectors which are necessary for electron, pion, kaon and proton separation. 
In particular, towards the electron end cap (negative $\eta$) it could be studied in the future, whether the tower depth could be increased up to 60~cm for higher $|\eta|$ to decrease the energy leakage for high energetic electrons, which are more probable in this region.
Here the projective design allows for such an extension at least for parts of the calorimeter. 

\paragraph{Barrel Hadronic Calorimeter: IHCAL \& OHCAL}\mbox{ }\\
\begin{figure*}[!t]
    \centering
    \includegraphics[width=0.55\linewidth]{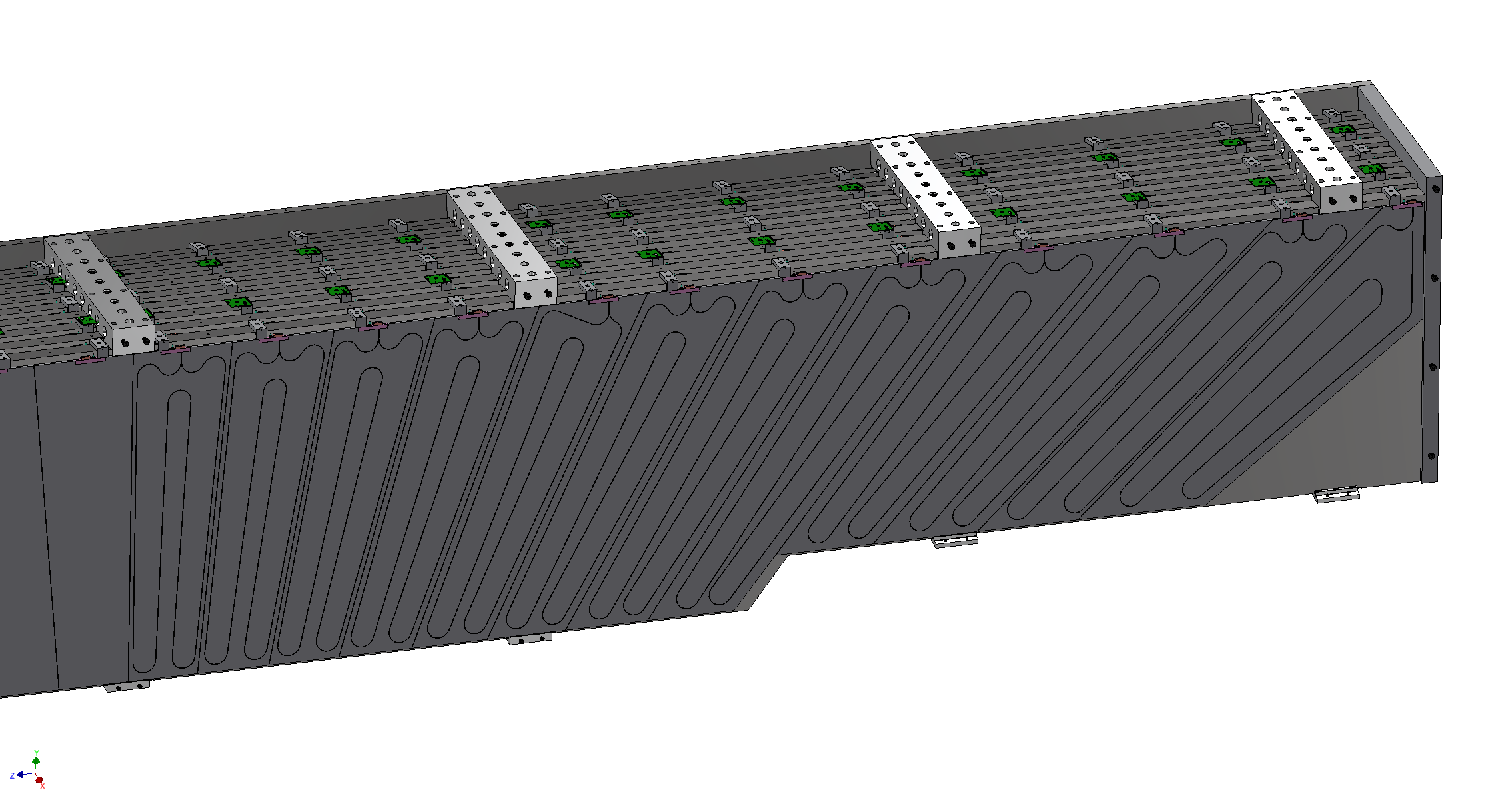}
    \includegraphics[width=0.35\linewidth]{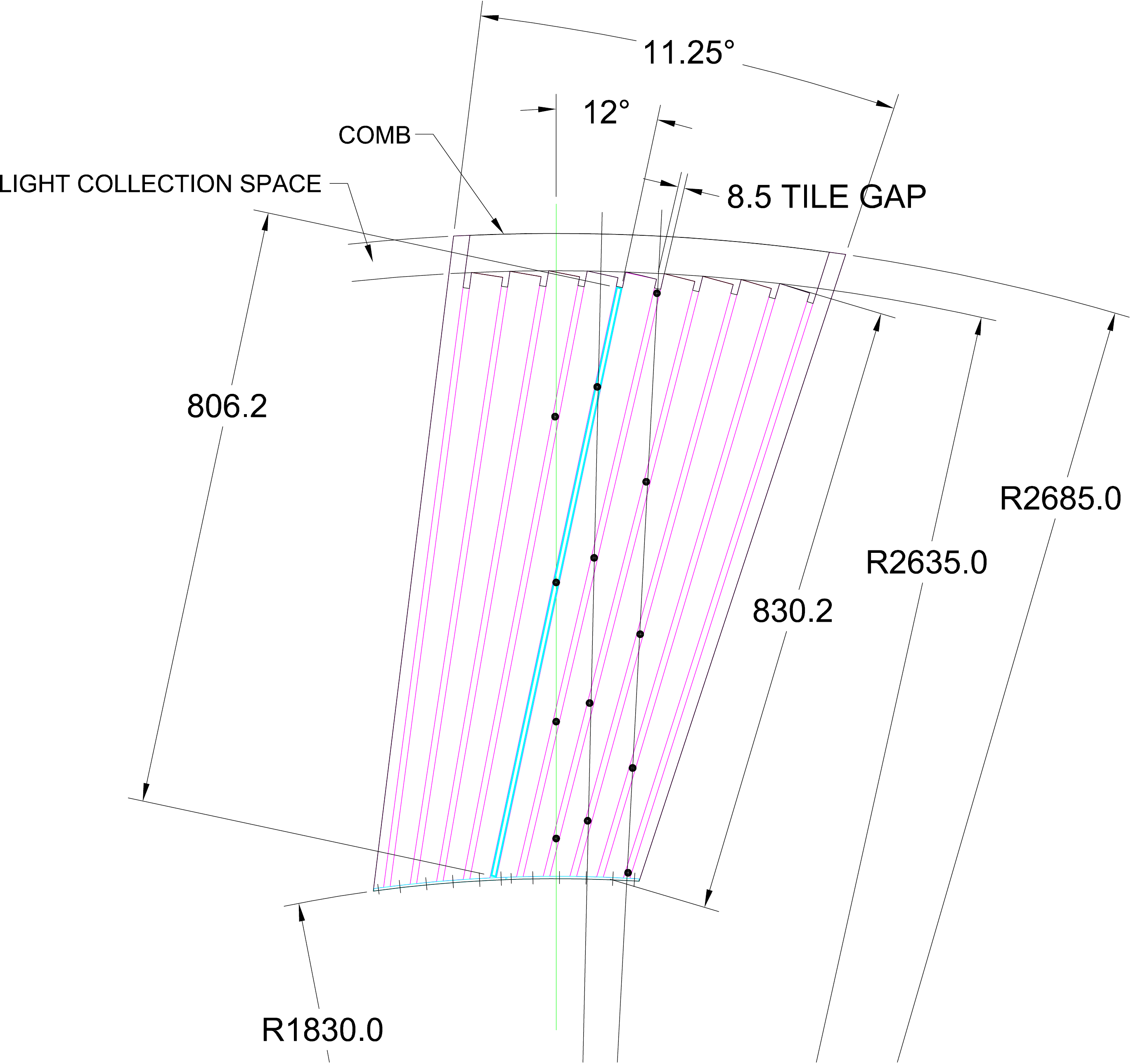}
    \caption{Left: Scintillator tiles in a layer of the OHCAL. 
    Right: Transverse cutaway view of an OHCAL module, showing the tilted tapered absorber plates.
    Light collection and cabling is on the outer radius at the top of the drawing.}
    \label{fig:inner_tile_pattern}
    \label{fig:hcal-6}
\end{figure*}
The Outer Hadronic Calorimeter (OHCAL) will be reused from the sPHENIX HCal~\cite{sPHENIX:2017lqb}, which instruments the large steel-based barrel flux return of the BABAR magnet.
The Inner Hadronic Calorimeter (IHCAL), as currently implemented in ECCE, is very similar in design to the sPHENIX inner HCAL in that it instruments the support for the barrel HCal to provide an additional longitudinal segment of hadronic calorimetry. 
The IHCAL provides useful data for overall calibration of the combined calorimeter system. \\
In the following, the construction of the scintillating tiles used in the outer and inner HCals is described, followed by a mechanical description of each calorimeter system.\\
\begin{table}[!t]
  \centering
  \begin{tabular}{lr}
    \toprule
    Parameter & Value \\
    \midrule
    Inner radius (envelope)     & 1820 mm\\
    Outer radius (envelope)     & 2700 mm\\
    Length (envelope)           & 6316 mm\\
    Material                       & 1020 steel \\ 
    \# towers in azimuth ($\Delta \varphi$)         & 64 \\
    \# tiles per tower                           & 5 \\
    \# towers in pseudorapidity ($\Delta \eta$)  & 24 \\
    \# electronic channels (towers)              & $64 \times 24 = 1536$ \\
    \# optical devices (SiPMs)                   & 5 $\times$ 1536 = 7680 \\
    \# modules (azimuthal slices)                & 32 \\
    \# towers per module  & $2 \times 24 = 48$ \\
    Total \# absorber plates  & $5 \times 64 = 320$ \\
    Tilt angle (relative to radius)   & 12$^{\circ}$ \\
    Absorber plate thickness at inner radius  & 10.2 mm\\
    Absorber plate thickness at outer radius  & 14.7 mm\\
    Gap thickness  & 8.5 mm\\
    Scintillator thickness  & 7 mm\\
    Module weight  & 12247 kg\\
    Sampling fraction  & 0.035\\
    Calorimeter depth & 4.0$\lambda/\lambda_0$ \\
    Moli\`ere radius $R_M$ for $\pi^\pm$ & 14.4 cm\\
    \bottomrule
  \end{tabular}
  \caption{Design parameters for the Outer Hadronic Calorimeter (OHCAL).}
  \label{tab:ohcalpar}
\end{table}
\begin{table}[!t]
    \centering
    \begin{tabular}{lr}
    \toprule
    Parameter & Value \\
    \midrule
    Inner radius (envelope)                       & 1350 mm\\
    Outer radius (envelope)                       & 1385 mm\\
    Material                                      & 310 stainless steel \\
    \# towers in azimuth ($\Delta \varphi$)             & 64 \\
    \# towers per module                             & $2 \times (12+15) = 56$ \\
    \# tiles per tower                               & 4 \\
    \# towers in pseudorapidity ($\eta > 0$)         & 24 \\
    \# towers in pseudorapidity ($\eta < 0$)         & 30 \\
    \# electronic channels (towers)                  & $64 \times 27 = 1728$ \\
    \# optical devices (SiPMs)                       & 4 $\times$ 1728 = 6912 \\
    Tilt angle (relative to radius)                   & 32 $^{\circ}$ \\
    Absorber plate thickness                          & 13 mm\\
    Gap thickness                                     & 8.5 mm\\
    Scintillator thickness                            & 7 mm\\
    \# modules (azimuthal slices)              & 32 \\
    Sampling fraction                                 & 0.059\\
    Calorimeter depth &  0.17$\lambda/\lambda_0$ \\
    \bottomrule
    \end{tabular}
    \caption{Design parameters for the Inner Hadronic Calorimeter (instrumented frame) for ECCE.}
    \label{tab:ihcalpar}
\end{table}
The basic calorimeter concept for the IHCAL/OHCAL is a sampling calorimeter with absorber plates tilted in the radial direction.
This design provides more uniform sampling in azimuth and provides information on the longitudinal shower development. 
The current design uses tapered plates for the OHCAL and non-tapered plates for the IHCAL.
Based on detailed studies, this design choice lowers the IHCAL machining cost without decreasing its performance.
Extruded tiles of plastic scintillator with an embedded wavelength shifting fiber are interspersed between the absorber plates and read out at the outer radius with SiPMs.
The tilt angle is chosen so that a radial track from the center of the interaction region traverses at least four scintillator tiles. 
Each tile is read out by a single SiPM, and the analog signal from each tile in a tower (five for the OHCAL, four for the IHCAL) are ganged to a single preamplifier channel to form a calorimeter tower. 
Tiles are divided in slices of $\eta$ so that the overall segmentation is $\Delta \eta \times \Delta \varphi \approx 0.1 \times 0.1$.\\
The scintillating tiles are similar to the design of the scintillators for the T2K experiment by the INR group (Troitzk, Russia) who designed and built 875~mm long scintillation tiles with a serpentine wavelength shifting fiber readout~\cite{Izmaylov:2009jq}.
Similar extruded scintillator tiles were also developed by the MINOS experiment. 
The properties of the HCal scintillating tiles and of the WLS fibers are detailed in Ref.~\cite{sPHENIX:2017lqb}.
The Kuraray single clad fiber is chosen due to its flexibility and longevity, which are critical in the geometry with multiple fiber bends.\\
The OHCAL is north-south symmetric and requires 24 tiles along the $\eta$ direction, whereas the IHCAL is asymmetric and has 12 towers in the forward direction and 15 towers in the backward direction. 
The OHCAL design therefore requires 12 different shapes of tiles for each longitudinal segment.
\Figure{fig:inner_tile_pattern} shows the tile and embedded fiber pattern for the OHCAL.\\
The major components of the OHCAL are tapered steel absorber plates and 7680 scintillating tiles which are read out with SiPMs along the outer radius of the detector.
The detector consists of 32 modules, which are wedge-shaped sectors containing 2 towers in $\varphi$ and 24 towers in $\eta$ equipped with SiPM sensors, preamplifiers, and cables carrying the differential output of the preamplifiers to the digitizer system on the floor and upper platform of the detector.
Each module comprises nine full-thickness absorber plates and two half-thickness absorber plates, so that as the modules are stacked, adjoining half-thickness absorber plates have the same thickness as the full-thickness absorber plates.
The tilt angle is chosen to be 12 degrees relative to the radius, corresponding to the geometry required for a ray from the vertex to cross four scintillator tiles.
Table~\ref{tab:ohcalpar} summarizes the major design parameters of the OHCAL, which are illustrated in \Figure{fig:hcal-6}.
Since the OHCAL will serve as the flux return of the solenoid, the absorber plates are single, long plates running along the field direction.
The IHCAL occupies a radial envelope bounded by a 50~mm clearance inside the solenoid cryostat and the outer radius of the BEMC.
The inner radius provides support for the BEMC and the HCal, while the end of the structure carries load to the OHCAL.\\
Table~\ref{tab:ihcalpar} shows the basic mechanical parameters of the IHCAL reference design.
The detector is designed to be built in 32 modules, which are wedge-shaped sectors comprising 8 gaps with 7 full-thickness plates and 2 half-thickness plates (so that as the modules are stacked, adjoining half-thickness plates have the same thickness as the full-thickness plates).
The modules contain 2 towers in $\varphi$ and 27 towers in $\eta$ equipped with SiPM sensors, preamplifiers, and cables carrying the differential output of the preamplifiers to the digitizer system on the floor and upper platform of the detector.
The instrumentation consists of 6912 scintillating tiles and optical devices, 1728 preamplifiers, and cabling.

\subsection{Hadron-End-Cap}
We envision the forward calorimeter system as an integrated ECal and HCal, where the installation units, where appropriate, are constructed in a common casing. 
These so-called modules consist of an electromagnetic calorimeter segment in the front which is part of the forward EMCal (FEMC) followed by a HCal segment which is part of the longitudinally separated HCal (LFHCAL). 
In between these segments a read-out section is foreseen for the ECal. 
The modules of up to 4 different sizes will be installed in half shells surrounding the beam pipe, which are movable on steel trolleys to give access to the inner detectors in the barrel in the hadron going direction. 
Each of these trolley should carry about 150 metric tons of weight.
This integrated ECal and HCal design reduces the dead material in the detector acceptance and allows for an easier installation in the experimental hall.
This implies that the construction of the modules has to happen in the same location to reduce shipping and assembly costs.
In the following, details on the FEMC will be discussed, followed by the design considerations and plans for the longitudinally separated HCal. \\
\begin{table}[t!]
  \centering
  \small
  \begin{tabular}{lll}
    \toprule
                      parameter &\textbf{FEMC}  &\textbf{LFHCAL}\\
        \midrule
        inner radius (envelope) & 17 cm  & 17 cm\\
        outer radius (envelope) & 170 cm & 270 cm\\
        $\eta$ acceptance       & $1.3 < \eta < 3.5$ & $1.2 < \eta < 3.5$\\
        tower information \\
        \hspace{1em} x, y ($R <$/$> 0.8$ m)     & 1 cm/ 1.65 cm & 5 cm\\
        \hspace{1em} z (active depth) & 37.5 cm & 140 cm\\
        \hspace{1em} z read-out       & 5 cm & 20 cm\\
        \hspace{1em} $\#$ scintillor plates   & 66 (0.4 cm each) & 70 (0.4 cm each)\\
        \hspace{1em} $\#$ aborber sheets          & 66 (0.16 cm Pb) & 60 (1.6 cm steel)\\
                                    &   & 10 (1.6 cm tungsten)\\
        \hspace{1em} weight                    & $\sim 6.4$ kg & $\sim 30.6$ kg\\
        \hspace{1em} radiation lengths & 18.5 $X/X_0$ & - \\
        \hspace{1em} interaction lengths & 1.0 $\lambda/\lambda_0$ & 6.9 $\lambda/\lambda_0$\\
        Moli\`ere radius $R_M$ & 5.2 cm (e$^\pm$ shower) & 21.1 cm ($\pi^\pm$ shower)\\
        Sampling fraction $f$ & 0.220 & 0.040\\
        $\#$ towers (inner/outer)         & 19,200/ 34,416 & 9040\\
        $\#$ read-out channels      & 53,616 & 7 x 9,040 = 63,280\\
    \bottomrule
  \end{tabular}
  \caption{Overview of the calorimeter design properties for the FEMC and the LFHCAL.}
  \label{tab:fwdcaloproperties}
  \centering
  \small
  \begin{tabular}{lll}
    \toprule   
        \textbf{Assembly Module Type} & \textbf{\# modules} \\
        \midrule
        8 LFHCAL tower modules (8M) & 1091 (total)\\
        \hspace{1em}    no FEMC towers in front & 538\\
        \hspace{1em}    200 FEMC towers (inner) & 87\\
        \hspace{1em}    72 FEMC towers (outer) & 466\\
        4 LFHCAL tower modules (4M)  & 76 (total)\\
        \hspace{1em}    no FEMC towers in front & 36\\
        \hspace{1em}    100 FEMC towers (inner) & 16\\
        \hspace{1em}    36 FEMC towers (outer) & 24\\
        2 LFHCAL tower modules (2M) & 2 (total)\\
        \hspace{1em}    50 FEMC towers (inner) & 2\\
        1 LFHCAL tower modules (1M) & 4 (total)\\
        \hspace{1em}    25 FEMC towers (inner) & 4\\
    \bottomrule
  \end{tabular}
  \caption{Number of assembly modules for the full combined FEMC and LFHCAL detector.}
  \label{tab:fwdcalomodules}
\end{table}
Both detector systems need to be able to handle the expected energies of incoming particles up to 150 GeV, based on simulated Pythia events for e+p collisions at 18x275 GeV$^2$. 
Due to the asymmetric collision system, these calorimeters are therefore focused strongly on high energetic particle shower containment while still providing good energy resolution down to lower energies.
\begin{figure*}[t]
    \centering
    \includegraphics[width=\linewidth]{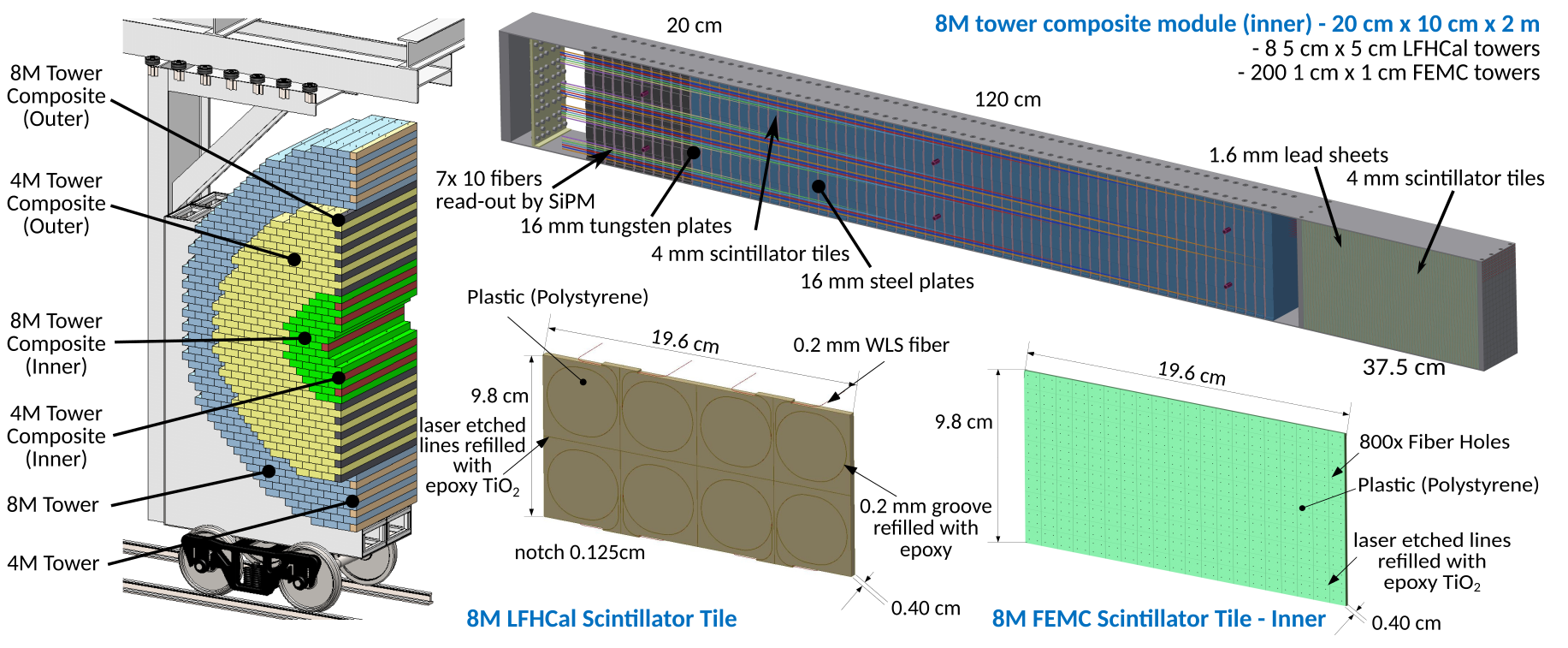}  
    \caption{Design pictures of the forward calorimeter assembly (left), 8-tower module design (top right) and single scintillator plates of the LFHCAL (bottom middle) and FEMC (bottom right) for an 8M tower module with embedded wavelength shifting fibers.}
    \label{fig:FCALdesign}
\end{figure*}

\paragraph{Hadron End-Cap Electromagnetic Calorimeter: FEMC}\mbox{ }\\
\label{sec:FEMC}
The forward ECal (FEMC) is a Pb-Scintillator shashlik calorimeter. 
It is placed at a distance of $z=3.07$ m from the interaction point in the hadron-going direction after the tracking and particle identification detectors. 
The detector is made up of two half disks with a radius of about 1.7m. 
The calorimeter is based on traditional Pb-Scint-Shashlik calorimeter designs like they have previously been used in ALICE, STAR and PHENIX. 
However, it employs more modern techniques for the readout and the scintillation tile separation. \\
Its towers have an active depth of 37.5~cm with additional space for the readout of about 5~cm.
Each tower consists of 66 layers of alternating 0.16~cm Pb sheets and 0.4~cm scintillator material, as listed in \Table{tab:fwdcaloproperties}.
Due to the high occupancy of the detector at large pseudorapities and the collimation of the particles in this area in physical space, the tower size varies depending on the radial position with respect to the beam axis.
Towers which are close to the beam pipe ($R < 0.8$~m) have an approximate tower size of $1 \times 1 \times 37.5$~cm$^3$. 
For the outer radii this granularity is not necessary and thus the size is increased to $1.65 \times 1.65 \times 37.5$~cm$^3$.
These numbers are intentionally well below the Moli\`ere radius of $R_m=5.18$~cm, thus showers will spread transversely over multiple towers.
In order to collect the light produced in the scintillator tiles, each scintillator and Pb-plate is pierced by four 0.2~mm diameter wavelength shifting fibers. 
These fibers are used to collect the light generated in the scintillators across all 66 layers. 
All four fibers are read out together by a single SiPM. \\
Multiple towers are contained in modules of either $20 \times 10$~cm$^2$ (8M), $10 \times 10$~cm$^2$ (4M), $5 \times 10$~cm$^2$ (2M) or $5 \times 5$~cm$^2$ (1M) in size. 
These module sizes match the 8-, 4-, 2- and 1-tower modules of the LFHCAL with which they share a 1.5~mm thin steel enclosure. 
Depending on the radial position, the FEMC packs 72 or 200 read-out towers in an 8M module. 
Due to the integration of the FEMC towers in the LFHCAL modules, the combined ECal and HCal modules are about 2.05 m long. 
A detailed drawing of the 8M inner scintillator tile design for the FEMC can be found in \Figure{fig:FCALdesign} (bottom right). 
The full 8M tile is made out of one piece.
In order to separate the light produced in different segments of the 8M-tile, the tile surface is subdivided into $1\times1$~cm$^2$ readout segments by CNC cutting or edging into the scintillator using a laser.
These 0.37~mm deep gaps (about 92\% of the tile thickness) are then refilled with a mixture of epoxy and Titanium-oxide (TiO$_{2}$) in order to reduce the light cross talk among different towers. 
The 4 fibers per tower are combined in a small light-collecting prism, which is directly attached to the SiPM with an effective photosensitive area of 9-16~mm$^2$ (ie. Hamamatsu S14160-3050HS). 
These SiPMs are most sensitive around wavelengths of 450~nm, thus the wave length shifting fibers have to be chosen accordingly to peak in a similar region.\\
The first signal processing happens after the ECal part of the module within the 5~cm space currently assigned for the FEMC read-out, realized using CMS HGCROC chips mounted on custom PCBs \cite{Thienpont:2020wau}, which can simultaneously process 72 channels. 
The signals are then transmitted via fiber optic cables to the end of the module for further processing.\\
A first full mechanical design for the joint LFHCAL and FEMC inner 8M module can be seen in \Figure{fig:FCALdesign}. 
Additionally, a first full illustration of a half shell is shown. 
The higher granular 8M and 4M FEMC-LFHCAL modules are indicated in green and red respectively, while the yellow and dark blue towers show the lower granularity 8M and 4M FEMC-LFHCAL modules. 
The lighter blue and orange modules reflect the modules only containing LFHCAL towers.\\
The majority of the FEMC is build of 8M modules, supplemented by 4M, 2M and 1M modules as outlined in \Table{tab:fwdcalomodules} to come closer to the beam pipe and allow for a vertical separation of the two half shells. 
The entire detector consists of approximately 53600 readout channels and provides a measurement of the energy of photons and electrons created in the collision going in the hadron-going (forward) direction. 

\paragraph{Hadron-End-Cap Hadronic Calorimeter: LFHCAL}\mbox{ }\\
\label{sec:LFHCAL}
The longitudinally separated forward HCal (LFHCAL) is a Steel-Tungsten-Scintillator calorimeter. 
The initial idea is based on the PSD calorimeter employed in the forward direction for the NA61/SHINE experiment \cite{Guber:109059}, but it has been extensively modified to meet the desired physics performance laid out in the Yellow Report. 
This longitudinally separated HCal is positioned after the tracking and PID detectors at $z=3.28$m from the center of the detector and is made up of two half disks with a radius of about 2.6m.  \\
The LFHCAL towers have an active depth of $\Delta z =1.4$ m with an additional space for the readout of about 20-30~cm depending on their radial position, as summarized in \Table{tab:fwdcaloproperties}. 
Each tower consists of 70 layers with alternating 1.6~cm absorber and 0.4~cm scintillator material and has transverse dimensions of $5 \times 5$~cm$^2$.
For the first 60 layers the absorber material is steel, while the last 10 layers serve as tail catcher and are thus made of tungsten to maximize the interaction length within the available space.  \\
In each scintillator, a loop of wavelength shifting fiber is embedded, as can be seen in \Figure{fig:FCALdesign} (bottom center). 
Ten consecutive fibers in a tower are read out together by a single SiPM, leading to 7 samples at different depth per tower. 
The towers are constructed in units of 8-, 4-, 2- and 1-tower modules to ease the construction and to reduce the dead space between the towers. 
Similar to the FEMC, the scintillator tiles in the larger modules are made out of one piece and then separated by gaps refilled with epoxy and Titanium oxide to reduce light cross-talk among the different readout towers. 
The wavelength shifting fibers running on the sides of the towers are grouped early on according to their readout unit and separated by thin plastic pieces over the full length.
The corresponding fiber bundles are indicated in \Figure{fig:FCALdesign} by different colors. 
They terminate in one common light collector, which is directly attached to a SiPM with an effective photosensitive area of 9-16~mm$^2$ (ie.\ Hamamatsu S14160-3050HS).
These 7 SiPMs per tower are then read out by a common readout design between the FEMC and LFHCAL based on the CMS HGCROC chips.
Alternatively, a common readout board which could be used for nearly all ECCE calorimeters is being pursued.
The entire detector consists of 63280 readout channels grouped in 9040 read-out towers and provides a measurement of the energy of hadronic particles created in the collision in the hadron-going (forward) direction.\\
The majority of the calorimeter is built of 8-tower modules ($\sim$1091) which are stacked in the support frame using a lego-like system for alignment and internal stability. 
The remaining module sizes are necessary to fill the gaps at the edges and around the beam pipe to allow for maximum coverage. 
The absorber plates in the modules are held to their frame by 4 screws each. To leave space for the read-out fibers, the steel and scintillator plates are not entirely square but equipped with 1.25~mm notches, creating the fiber channels on the sides, as can be seen in \Figure{fig:FCALdesign} (bottom center) for a scintillator plate. 
In order to protect the fragile fibers, the notched fiber channels are covered by 0.5~mm thin steel plates after module installation and testing.
For internal alignment we rely on the usage of 1-2~cm steel pins in the LFHCAL part which are directly anchored to the steel or tungsten absorber plates. 
Consequently, the modules are self-supporting within the outer support frame. 
The support frame for the half disks is arranged on rails which allows the HCal and ECal to slide out to the sides and gives access to the inner detectors. 
In addition, the steel in the LFHCAL serves as flux return for the central 1.5T BABAR magnet.
As a consequence, a significant force is exerted on the calorimeter, which needs to be compensated for by the frame and internal support structure.  

\section{Calorimeter Performance}
\label{sec:performance}
The ECCE electromagnetic and hadronic calorimeters are designed to meet the criteria outlined in the Yellow Report.
In the following, the expected performance of the different systems is presented based on standalone and full detector GEANT4 simulations.
\begin{figure}[t]
    \includegraphics[width=\linewidth]{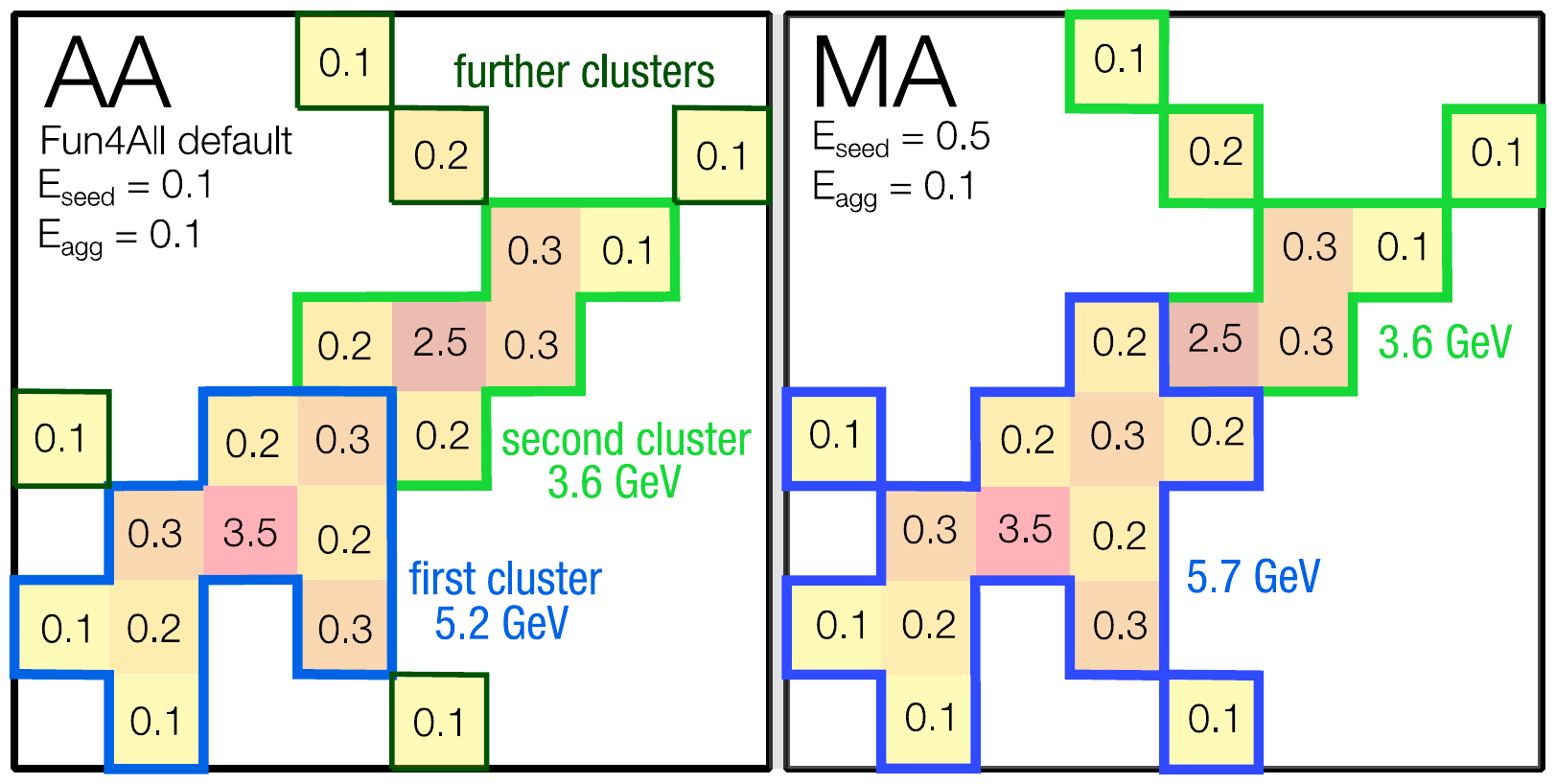}
    \caption{Clusterization algorithms visualized on an example energy deposit from two particles ($E_1^\mathrm{true}=6$ GeV and $E_2^\mathrm{true}=4$ GeV) in calorimeter towers of arbitrary size. 
    Presented are the AA (Aggregate-All) and the MA (Modified-Aggregation) clusterizers.
    The found clusters are outlined in color and their reconstructed energy is indicated in the figure. 
    The same seed and aggregation energy thresholds are assumed for both algorithms in this example. }
    \label{fig:clusterizersEIC}
\end{figure}
    
\subsection{Clusterization}
    The energy deposit from an electromagnetic or hadronic shower is generally spread over multiple towers.
    The magnitude of this effect depends on the tower size relative to the Moli\`ere radius ($R_M$) of the material used and is more prominent for hadronic showers.
    The Moli\`ere radius is defined as the radius in which 90\% of the shower energy is contained, where electron-induced showers are used for the ECals and charged pion-induced showers are used for the HCals.
    Since $R_M$ is in all cases larger than the individual tower sizes in the different calorimeters, it becomes apparent that the full shower can only be reconstructed when the information of multiple towers is combined.
    Different reconstruction algorithms can be employed in order to group towers containing energy deposits into so-called clusters, which are the main objects used in physics analyses.
    The performance of these algorithms mostly depends on the calorimeter occupancy for a given event. 
    While showers from single electromagnetic particles are mostly trivial to reconstruct, a significant challenge is posed by overlapping particle showers, for example in a jet or from high energetic neutral meson decays. 
    In the latter case, the decay photon showers, e.g. from $\pi^0\rightarrow\gamma\gamma$, can not be separated within the calorimeter granularity above a certain particle energy due to the decay kinematics.
    Thus, extensive studies were performed to increase the separation power between single and multi-particle showers and to absorb as much of the deposited energy as possible during the so-called clusterization procedure.
    This procedure always starts with the highest energetic tower in the calorimeter, which is required to contain an energy deposit above a seed energy threshold ($E_\mathrm{seed}$).
    \begin{figure}[t!]
      \centering
      \includegraphics[width=0.48\textwidth]{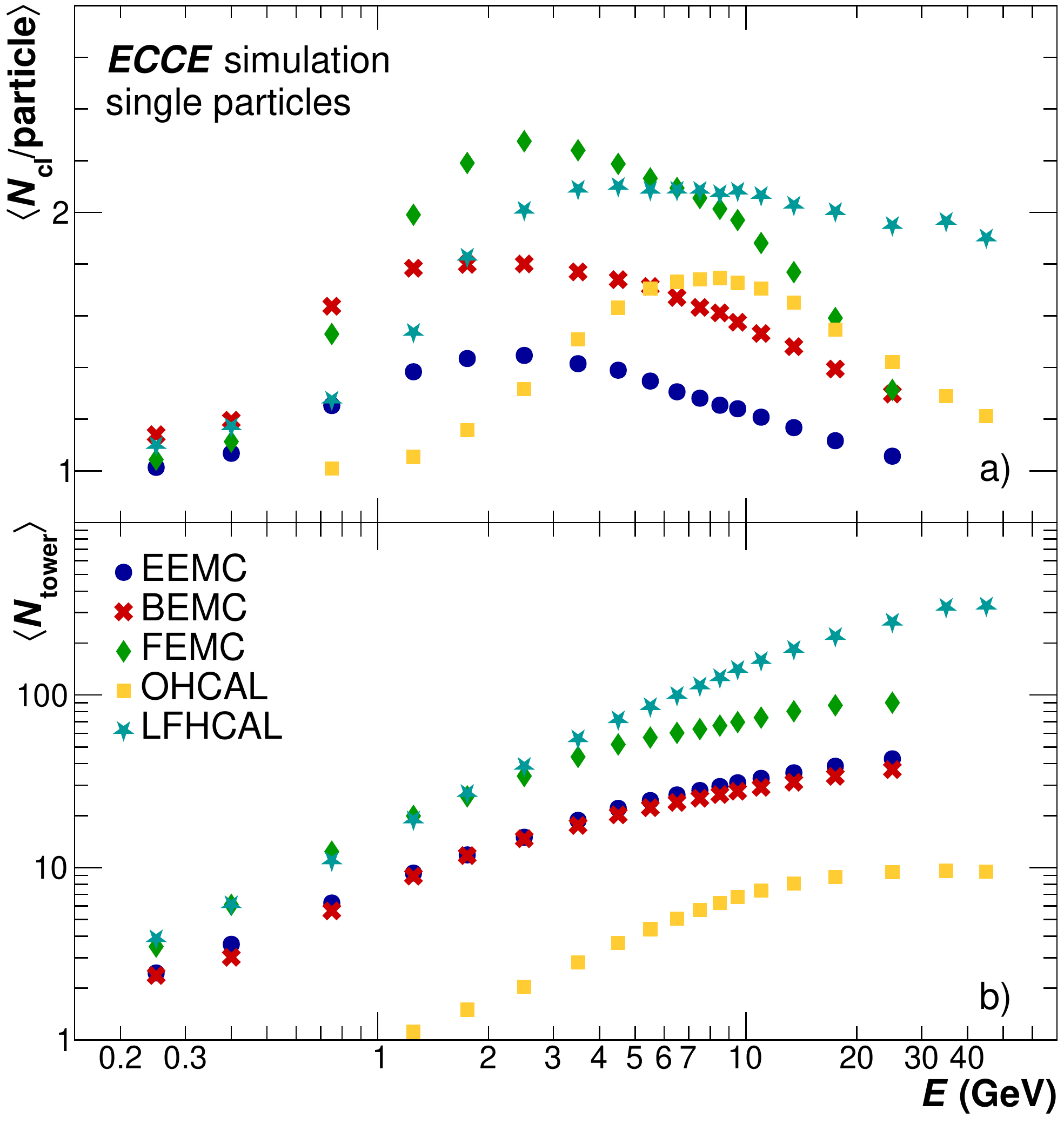}
      \caption{Mean number of clusters per generated particle (a) and average number of towers aggregated within a cluster (b)  as a function of generated particle energy using the MA clusterizer for the different ECCE calorimeters. }
      \label{fig:ntowandncluspartFEMC}
    \end{figure}
    \begin{figure*}[t!]
      \centering
      \includegraphics[width=\textwidth]{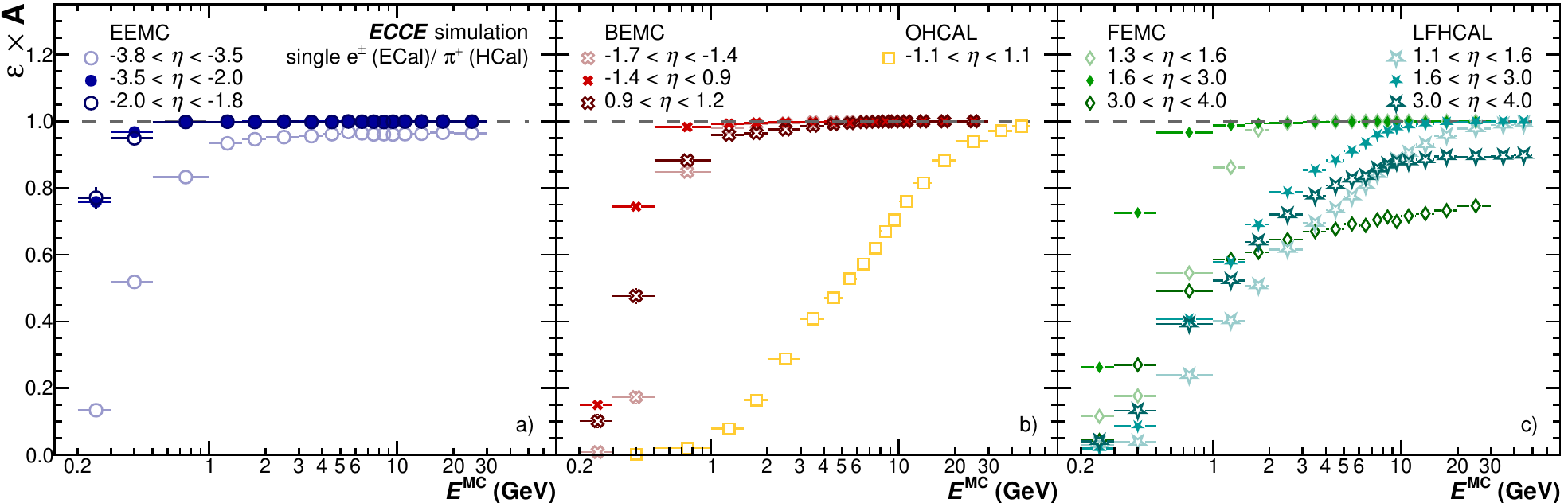}\\
      \caption{Cluster reconstruction efficiencies in the EEMC, BEMC and FEMC for electrons (a--c) and in the OHCAL and LFHCAL for charged pions (b and c) reconstructed with the MA-Clusterizer. 
      The efficiencies are calculated according to Equation~\ref{eq:caloeffi}.}
      \label{fig:clusterreceffiEBF}
    \label{fig:clusterreceffiHCLF}
    \end{figure*}
    Additional neighboring towers are added to the cluster if their energy exceeds a certain aggregation threshold ($E_\mathrm{agg}$).
    The thresholds ($E_\mathrm{seed}$ and $E_\mathrm{agg}$) for the different ECals and HCal have been optimized to reduce false seeding from minimum ionizing particles and to suppress noise during aggregation.
    Their values are tower size and calorimeter type dependent, with approximate values of $E_\mathrm{seed}^\mathrm{ECal} = 100$ MeV and  $E_\mathrm{agg}^\mathrm{ECal}=$ 5--10 MeV or $E_\mathrm{seed}^\mathrm{HCal} =$ 100--500 MeV and $E_\mathrm{agg}^\mathrm{HCal}=$ 5--100 MeV for the ECals or HCals, respectively.
    Two main algorithms have been explored for the cluster reconstruction: the so-called aggregate-all (AA) and modified-aggregation (MA) clusterizers. 
    The AA clusterizer associates all towers sharing a common side with already aggregated towers in the cluster and only stops the aggregation when no further tower above $E_\mathrm{agg}$ can be found.
    At this point, the already aggregated towers are removed from the sample and a new seeding starting from the next highest energetic tower is performed.
    Since this approach can aggregate energy deposits from multiple particles depending on the occupancy, a subsequent splitting of the cluster should be performed based on the number of maxima found in the energy distribution.
    This cluster splitting procedure is necessary when AA clusters are meant to be used for single particle analyses.
    The MA clusterizer works similar to the AA clusterizer, however the algorithm stops when a neighboring tower with larger energy than the already aggregated tower is found.
    It also aggregates towers that share a common corner, thus a $3 \times 3$ tower window around each already aggregated tower is inspected.
    This algorithm is preferred for the reconstruction of hadronic showers in high granularity calorimeters, since the energy deposits can fragment over a large amount of towers.
    \Figure{fig:clusterizersEIC} shows these algorithms applied to an example energy deposit in a calorimeter, where different clusters are reconstructed based on the various aggregation conditions.
    The MA clusterizer is also the only clusterizer employed in the LFHCAL cluster reconstruction due to the additional z-segmentation of the calorimeter.
    For this, the MA clusterizer also allows the inclusion of neighboring towers in z-direction sharing an edge or corner with already aggregated towers.\\
    As can be seen in \Fig{fig:ntowandncluspartFEMC}, the various ECCE calorimeters show visible differences in the average number of towers they aggregate per MA-based cluster.
    In addition, the top panel of the same figure shows the mean number of clusters per generated particle, which is approximately one at low energies for the MA clusterizer, but reaches up to two clusters per particle at higher energies for the LFHCAL and FEMC.\\    
    Overall it was found that the MA clusterizer performs slightly better than the AA clusterizer for all calorimeters and especially in events with a higher occupancy in the different detectors.
    The MA clusterizer is therefore chosen as default for the following detector performance studies.\\
    An important property of any clusterization algorithm and calorimeter is the efficiency with which a cluster can be reconstructed for any given particle.
    \Fig{fig:clusterreceffiEBF} shows the reconstruction efficiencies for electrons and charged pions for the different ECals and HCal as a function of generated particle pseudorapidity calculated according to Equation~\ref{eq:caloeffi}.
    \begin{equation}
    \epsilon \cdot a= \frac{N_{\mathrm{clus,Ntow}>1} \text{ in calorimeter}}{N_\text{MC gen. particles} \text{ in acceptance}}
    \label{eq:caloeffi}
    \end{equation}
    In the calculation, only clusters formed according to the seeding and aggregation thresholds are used and additionally required to be made of more than a single tower.
    Furthermore, only one cluster per generated particle is considered for the calculation of $\epsilon \cdot a$ to avoid counting multiple clusters of a single particle (e.g. due to an induced pre-shower).
    The latter requirement is necessary to reject secondary low energy clusters or showers that are not contained in the calorimeter (e.g.\ on the outer and/or inner edges).
    The efficiencies show that at low energies, the seed and aggregation thresholds decrease the reconstruction efficiency.
    Moreover, an edge effect at low and high pseudorapidity (e.g. strong shower leakage) can be observed even at higher energies also reflecting a small remaining effect from acceptance losses in the calorimeters mostly due to deflection of the original particles in the inner detectors. 
    Additionally, the non-uniform $\varphi$-coverage of the LFHCAL and FEMC is clearly visible, yielding a lower average efficiency for $\eta > 3$. 
    \begin{figure*}[!t]
        \centering
        \includegraphics[width=\textwidth]{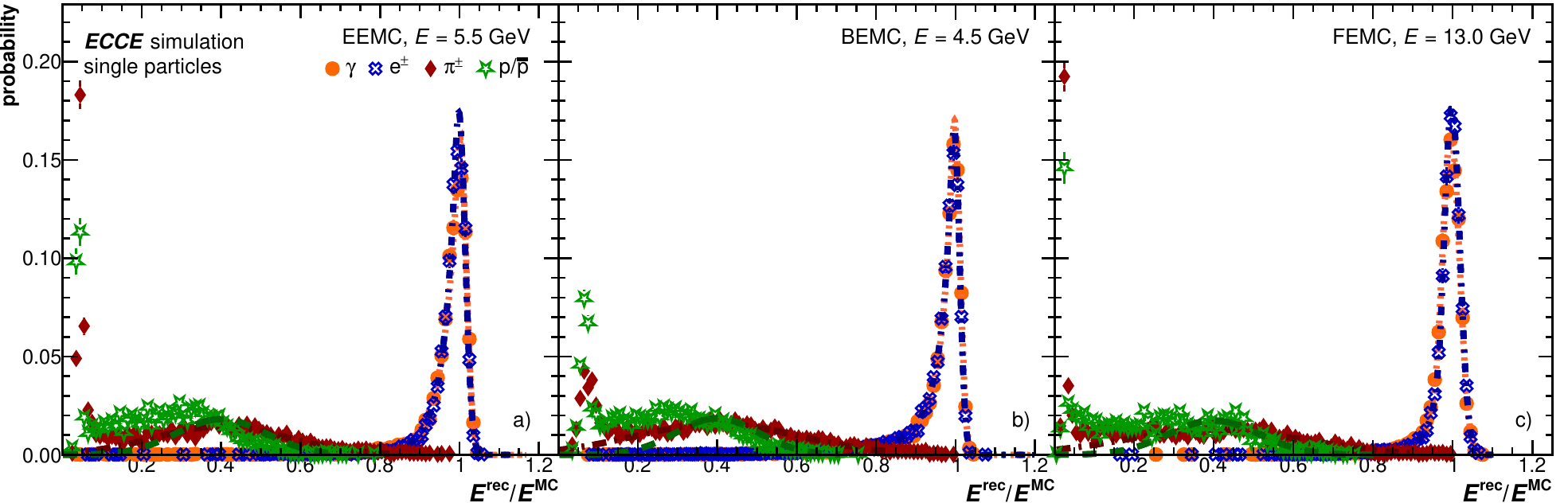}\\ 
        \includegraphics[width=0.66\textwidth]{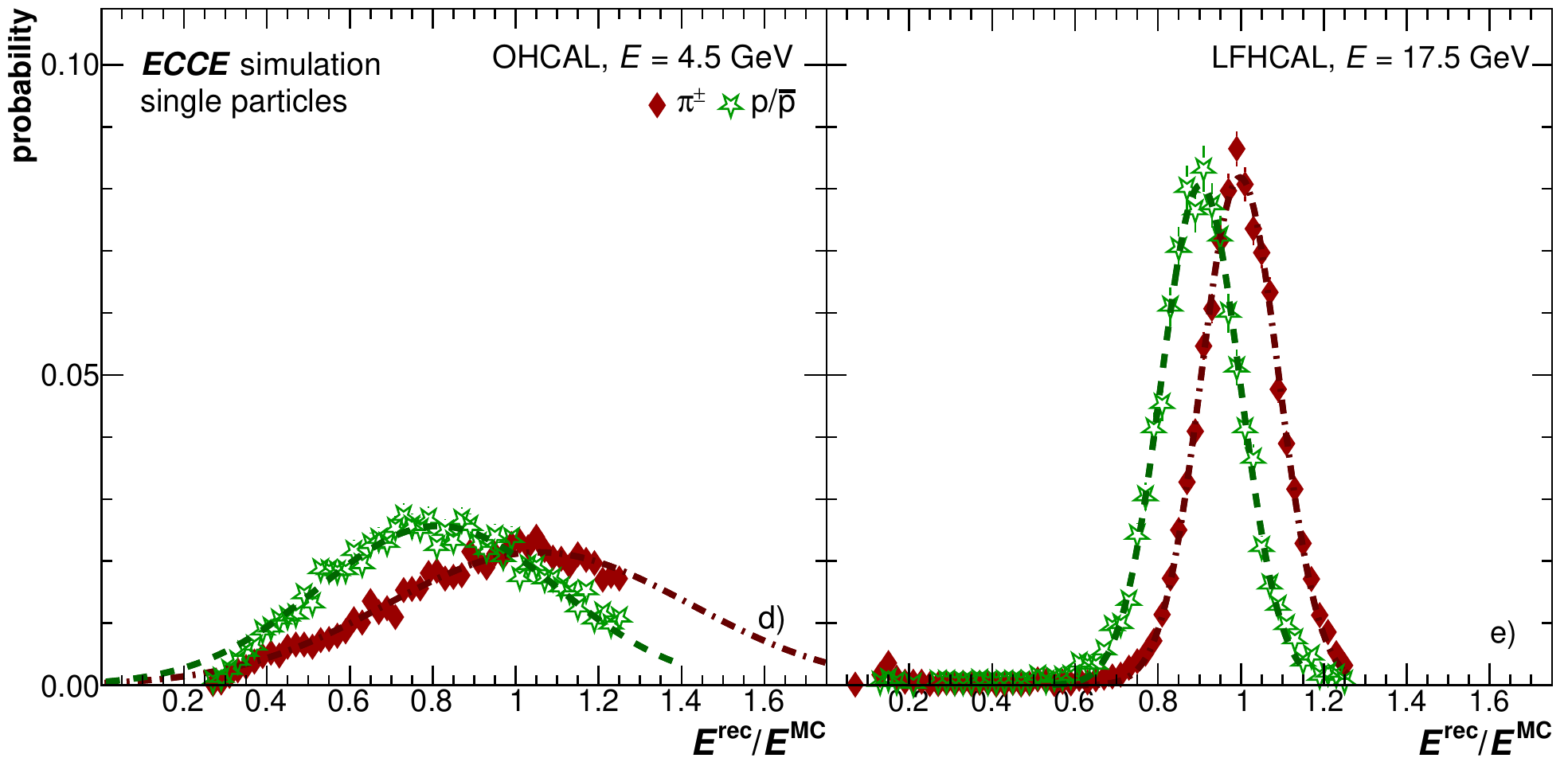} 
        \caption{Energy resolution for different particles generated in single particle simulations at fixed energies as measured by the electromagnetic calorimeters EEMC, BEMC, and FEMC (a--c) and the hadronic calorimeters OHCAL and LFHCAL (d and e). }
         \label{fig:eresodistributioneemcbecalfemc}
   \end{figure*}

\subsection{Energy resolution}
  The energy resolution for the ECals and HCals is evaluated based on single particle simulations for photons, electrons, pions and protons generated for $0.2 < E < 30 (50)$ GeV. 
  For these studies the reconstructed energy deposits in the towers are combined into clusters using the MA clusterizer with the aforementioned seed and aggregation energy settings for each calorimeter. 
  The energy scale of the calorimeters is calibrated such that in simulations without material in front of the calorimeter the reconstructed electron energy over the generated energy is approximately unity. 
  Thus, this calibration corrects the ECals and HCal to approximately the same energy scale.
  No $\eta$ dependent corrections for the energy response are introduced so far.
  \begin{figure}[t!]
      \centering
      \includegraphics[width=0.4\textwidth]{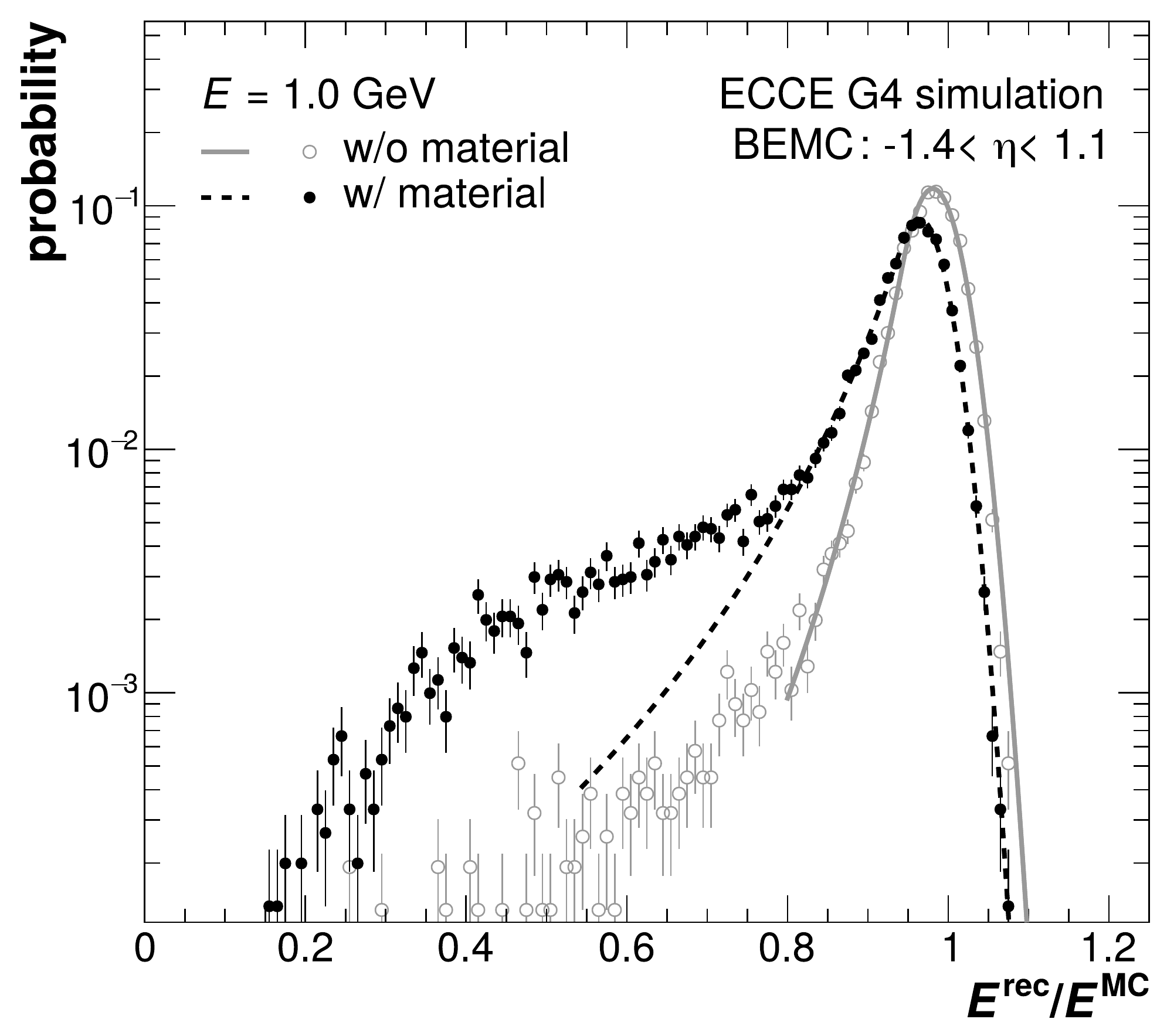}
      \vspace{-0.2cm}
      \caption{Comparison of the energy resolution for electrons generated in single particle simulations at $E= 1$ GeV (top) and $E= 8$ GeV (bottom) as measured by the BEMC (left) and FEMC (right) without additional material in front of the calorimeter and in the full detector setup.}
      \label{fig:materialresopeak}
  \end{figure}
  \begin{figure}[t!]
      \centering
      \includegraphics[width=0.45\textwidth]{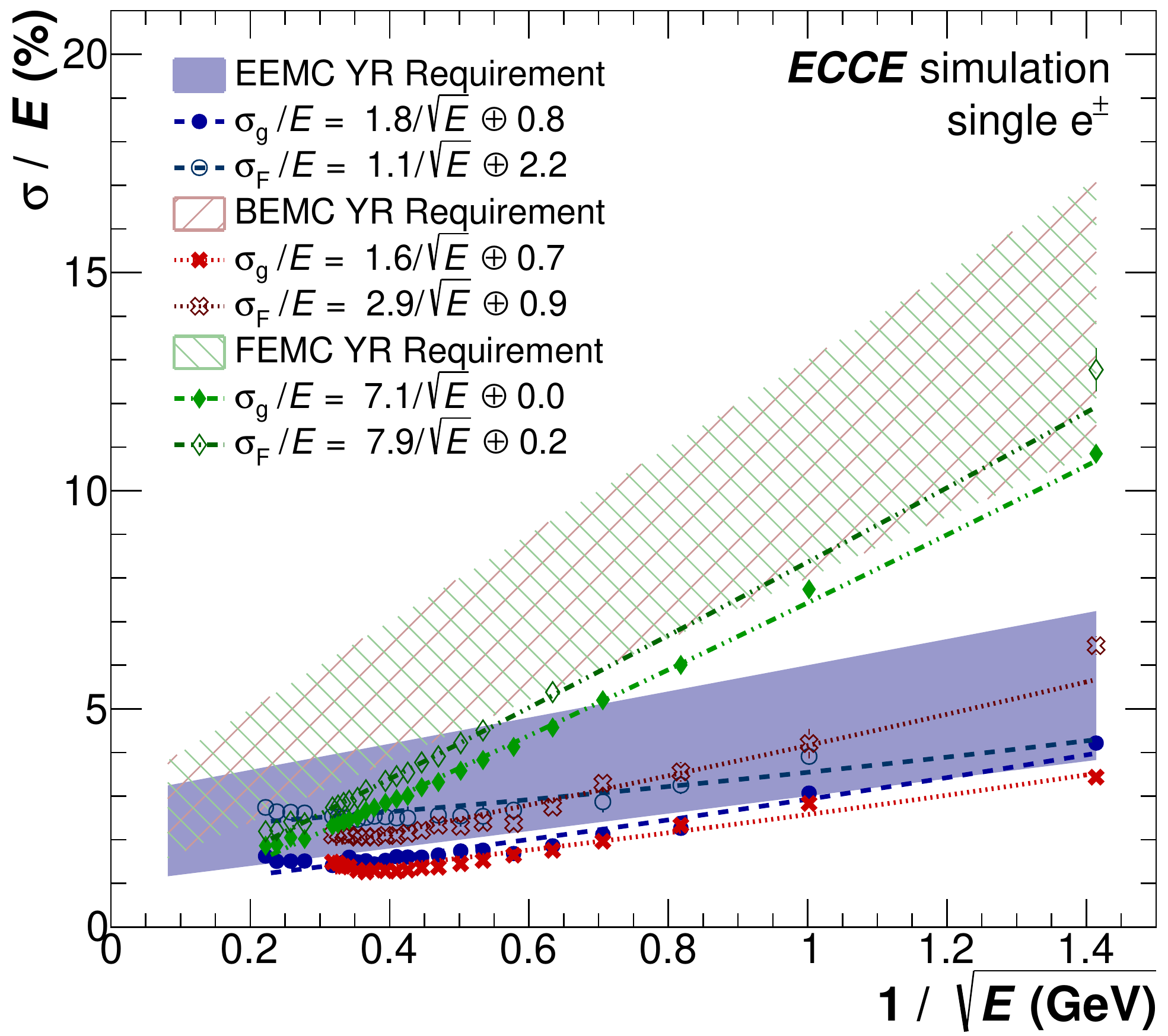}\\
      \includegraphics[width=0.45\textwidth]{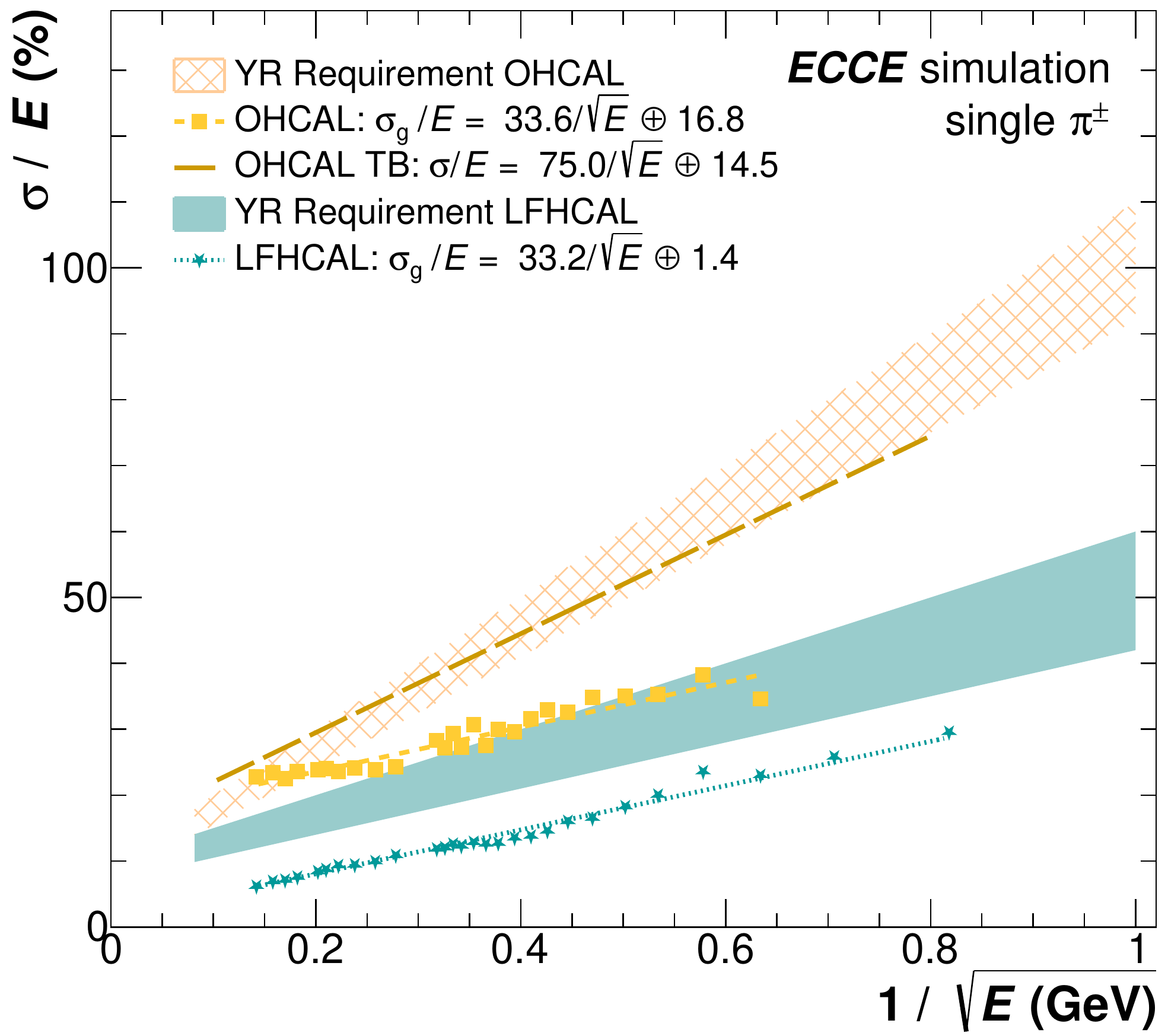}
      \caption{Energy resolution for electrons (charged pions) generated in single particle simulations with energies between 0.2 and 20 (50) GeV as measured by the different ECals (top) and the different HCals (bottom) in the central acceptance of the corresponding detectors.
      The shaded bands show the requirements as extracted from the yellow report for the different calorimeters. 
      The data points and fits indicated as $\sigma_g/E$ are based on the Gaussian width of the resolution peaks, while $\sigma_F/E$ is based on the FWHM.
      The energy resolution based on a test beam for the OHCAL is shown for comparison \cite{sPHENIX:2017lqb}.}
      \label{fig:eresocalos}
  \end{figure}
  \Figure{fig:eresodistributioneemcbecalfemc} shows the energy response $E^\mathrm{rec}/E^\mathrm{MC}$ for the various particle species and in each calorimeter.
  By construction, the electron and photon response in the ECals peaks around unity with a strong excess that is accompanied by a visible tail towards lower values. 
  This tail is a result of multiple effects.
  First, the clusterization in the calorimeter is not perfect (see clusterization chapter) and thus not all energy of an incoming particle is reconstructed.
  In addition, for these studies only the highest energetic cluster in each event is selected, which combined with the clusterizer performance leads to a smearing to lower $E^\mathrm{rec}/E$ values.
  Further smearing comes from bremsstrahlung losses of the electrons in the magnetic field as well as from material interaction of photons that could lead to photon conversions, as seen in \Fig{fig:materialresopeak}.
  The figure shows a comparison of the energy response for the BEMC with and without the remaining ECCE detector material in front, highlighting an increasing tail at lower $E^\mathrm{rec}/E$ due to the additional material.
  In the following studies, contributions from photon conversions are not rejected and thus are still contained in the photon sample.
  The left side tails of the resolution peaks can also arise through particles hitting the support material in between the towers.
  The reconstructed energy loss from hitting and subsequently channeling in the passive support structures is a major factor to be considered for the calorimeter design.
  Initial studies have shown that already a 2 mm carbon fiber support structure between the EEMC towers is enough to significantly deteriorate the energy resolution.
  As such, the supports were optimized to the current design of 0.5 mm carbon sheets, which greatly improves the energy resolution.
  Further improvements are possible with carbon support grids holding multiple crystals that are further separated by a thin foil.
  Similar support material considerations are to be made for the BEMC, where the current design employs 2 mm carbon fiber sheets.
  Charged hadrons deposit in the majority of cases only a minimum ionizing signal in the ECals, which is visible as a strong peak at low $E^\mathrm{rec}/E$ values.
  However, there is also a non-negligible amount of charged hadrons that deposit 40\% or more of their energy in the ECals, which can negatively impact the HCal energy resolution.
  For the HCals, the charged pions and protons peak around unity, whereas remaining shower leakage from electron showers out of the ECals is mostly negligible.
  \Figure{fig:eresodistributioneemcbecalfemc} also highlights a shifted peak for protons compared to pions in the HCals which can be explained by a loss of visible energy for baryons.
  In future studies, this effect could be counteracted for the LFHCAL by shower depth analyses and subsequent application of a correction factor for the loss of visible energy.\\
  In order to determine the energy resolutions of the different calorimeters, the $E^\mathrm{rec}/E$ distributions are fitted with crystalball functions in order to determine the peak width.
  This width can either be taken from the Gaussian component or from the full width at half maximum (FWHM).
  The slightly larger values of the latter are a reflection of the asymmetric $E^\mathrm{rec}/E$ distribution as described above.
  Based on the fit values, \Fig{fig:eresocalos} shows the energy resolution for electrons in their generated energy range in the ECals and for charged pions in the HCals.\\
  All ECal resolutions, based either on the Gaussian sigma ($\sigma_\mathrm{g}$) or the FWHM ($\sigma_\mathrm{F}$), are well within the limits imposed by the YR and even exceed the requirements in the case of the BEMC by a significant amount.
  Thus, despite the smeared $E^\mathrm{rec}/E$ peaks from the full ECCE detector simulation, the resolution is still within the imposed limits.
  In addition, a minimal pseudorapidity dependence for all calorimeters is observed, but none of the $\eta$-regions fail to deliver the required YR performance.\\    
  For the HCals, I/OHCAL and LFHCAL, a similar behavior is observed, where the resolutions are found to be better than the YR requirements with $\sigma/E=(31-34\%)/\sqrt{E}\oplus(17-18\%)$ and $\sigma/E=(33-44\%)/\sqrt{E}\oplus(1.4\%)$ by about 1--20\% and constant 8\%, respectively.
  This also holds true for both tested particle species ($\pi^\pm$ and protons) and in each $\eta$ region individually.
  In addition, the HCal resolution is compared to the sPHENIX test beam data and shows a better resolution in the presented simulations \cite{sPHENIX:2017lqb}, which can be explained by an imperfect simulation setup of the details of the calorimeter response.
  
\subsection{Position resolution}
\label{sec:PositionResolution}
  A significant fraction of physics observables either directly or indirectly require a good position resolution of the reconstructed clusters in the calorimeters. 
  For example, the jet reconstruction clusters objects which are reconstructed in a given radial cone and thus position inaccuracies especially in difficult pseudorapdity regions can deteriorate the physics performance. 
  Moreover, charged particle association or neutral cluster determination via track matching (see next section) depends on the cluster position resolution as much as on the tracking resolution.
  \begin{figure}[t!]
    \centering
    \includegraphics[width=\linewidth]{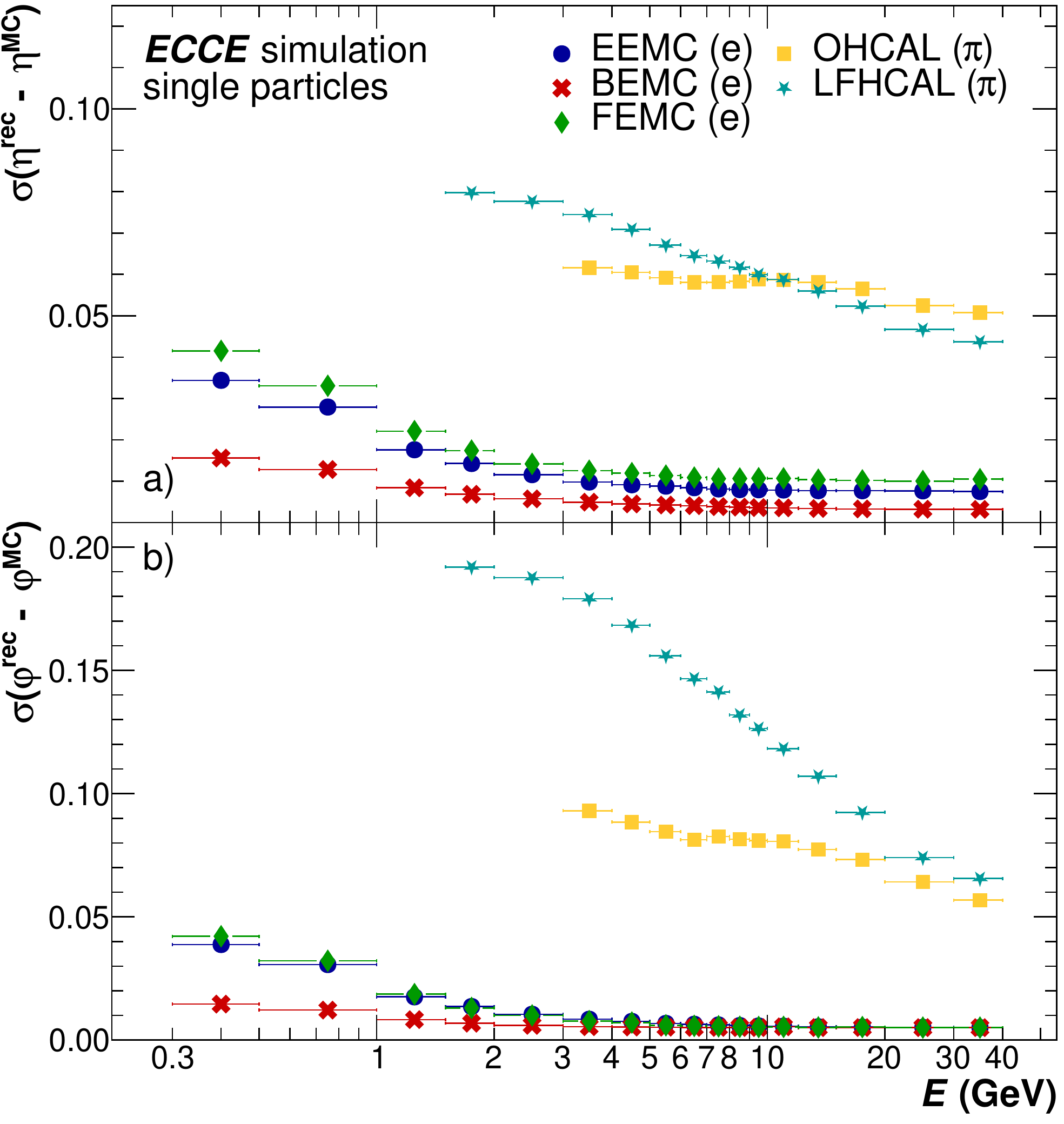} 
    \caption{ Position resolution in $\eta$ (top) and $\varphi$ (bottom) for electrons or charged pions generated in single particle simulations with energies between 0.2 and 40 GeV as measured by the different calorimeters in the central acceptance of the corresponding destectors without a magnetic field. }
    \label{fig:posreso}
  \end{figure}

  \begin{figure*}[t!]
    \centering
    \includegraphics[width=\textwidth]{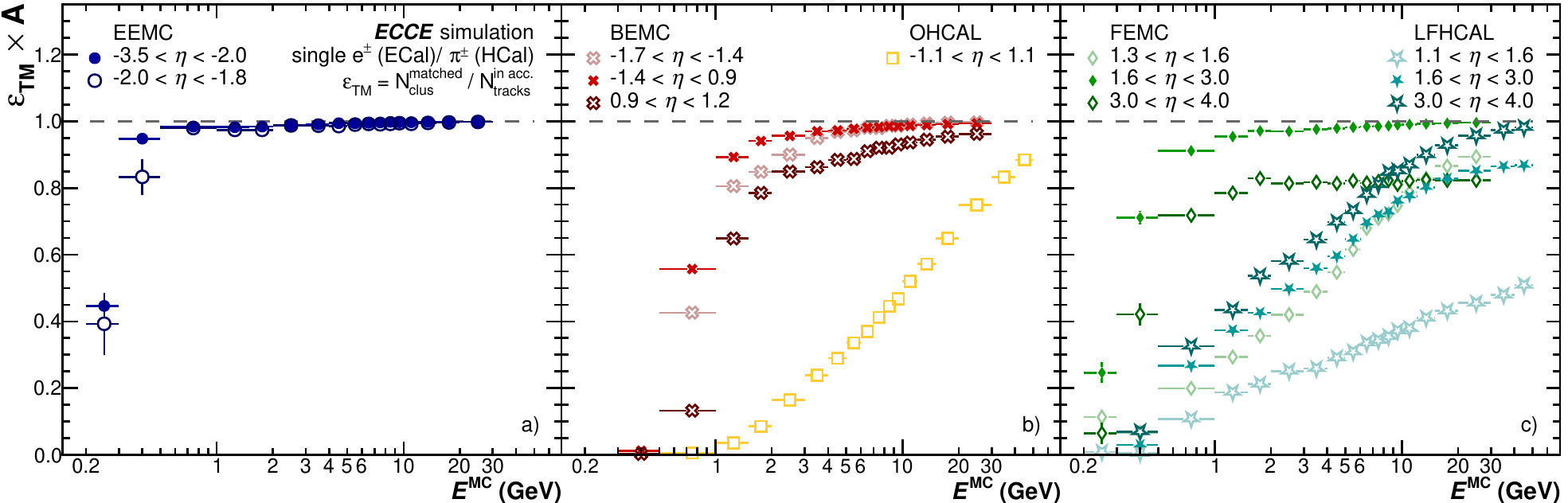}
    \caption{Track matching efficiencies for electrons reconstructed with the MA-Clusterizer in the EEMC (a), the BEMC (b) and the FEMC (c) and for charged pions reconstructed in the OHCAL (b) and the LFHCAL (c). The efficiencies are relative to the number of reconstructed tracks according to \Equation{eq:tmeffitrk}. }
    \label{fig:tmeffiEBF}
  \end{figure*}
  To determine the pure position resolution of the clusterization algorithm and intrinsic calorimeter granularity single particle simulations without a magnetic field have been used.
  This setup allows to separate between the intrinsic position resolution in the respective calorimeters and effects arrising from a larger inclination angle at the calorimeter surface as well as inaccuracies in the particle propagation through the material due to the 1.4T magnetic field.
  For the track-to-cluster matching under realisitic conditions within a magnetic field the $\eta$ and $\varphi$ coordinates for charged particles are calculated by propating the tracks through the detector material to approximately half the depth of each calorimeter.
  The median cluster depth is however $\eta$ dependent for non projective calorimeters.
  Consequently, the mean shift in the $\eta$-position has to be corrected for the forward and backward calorimeters based on the zero-field data.
  \Figure{fig:posreso} presents the width of the difference of the generated particle $\eta$($\varphi$) and the reconstructed cluster position in $\eta$($\varphi$) in the different calorimeters.
  For all electro-magnetic calorimeters an excellent resolution of about $0.01-0.015$ in pseudorapidity is observed which only degrades slightly towards lower energies.
  The $\varphi$-resolution for highly energetic particles is similarly good with $\Delta \varphi = 0.02$ (corresponding to $1.15$ degrees). 
  It is mainly determined by the size of the single towers in $\Delta \varphi$ of the respective calorimeter and the width of the electro magnetic shower.
  Due to the larger tower sizes and wider spread of hadronic showers without a very well defined core the $\eta$ and $\varphi$ resolutions of the hadronic calorimeters are slightly worse in both dimensions. 
  The resolutions for the LFHCAL could be further improved in the future by taking into account the correct depth of the shower as well, which so far has not been considered in the position calculation.
\subsection{Track-Cluster matching}
\begin{figure}[!ht]
    \centering
    \includegraphics[width=0.49\textwidth]{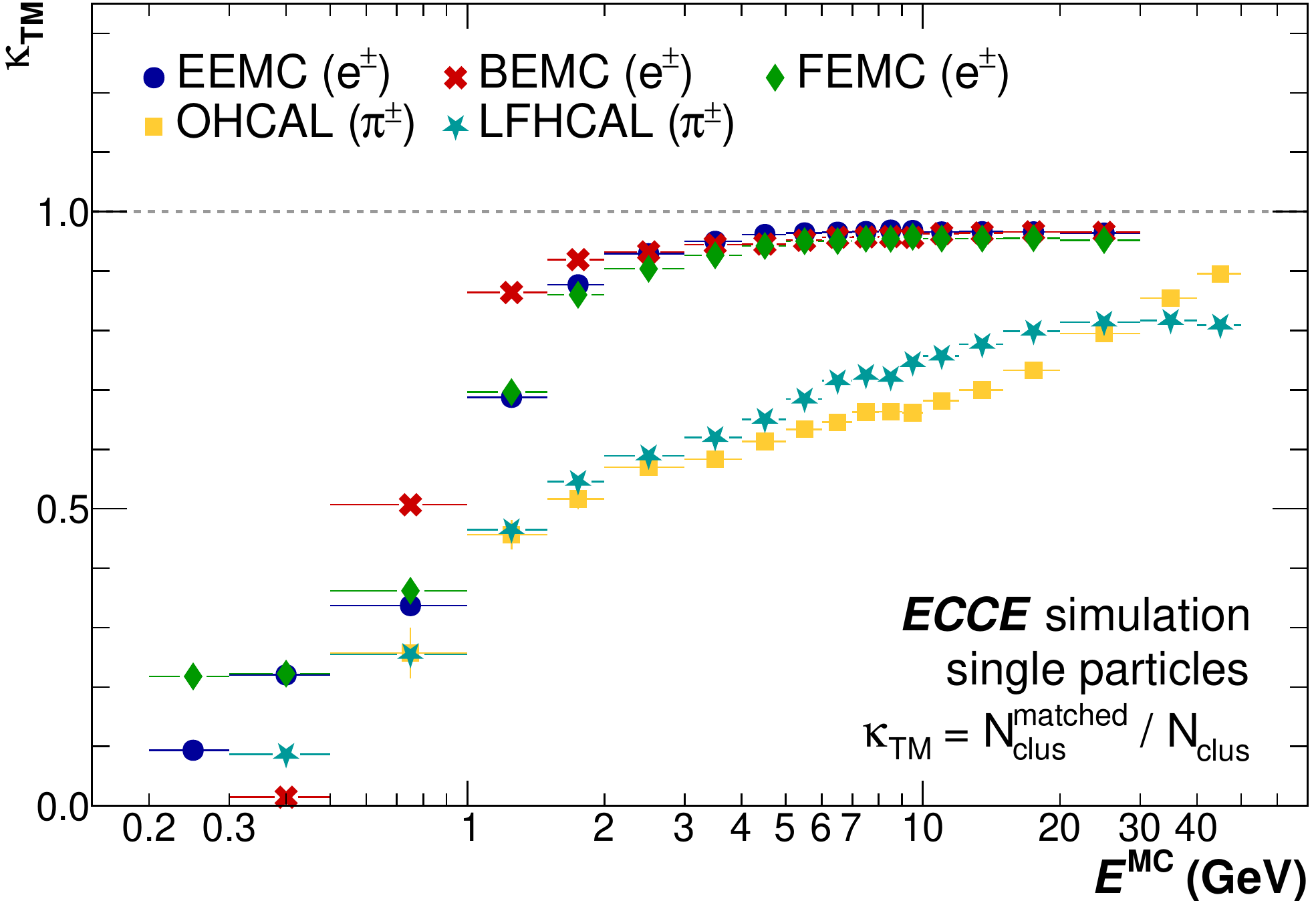}
    \caption{Track matching efficiencies calculated relative to the number of reconstructed clusters ($\kappa_\mathrm{TM}$). Single particle simulations of electrons in the detector acceptance of the ECals and of charged pions in the acceptance of the HCals are used to obtain the efficiencies.}
    \label{fig:tmeffitrkclus}
  \end{figure}

  The position resolution described in the previous chapter is a necessary ingredient for performance studies of the cluster-to-track matching.
  This matching is needed for particle identification studies, like electron selection via charged pion rejection or cluster neutralization for photon analyses.
  Moreover, the track matching procedure is a crucial ingredient for particle flow-based jet measurements.\\
  The track matching efficiency can be calculated as the number of track-matched clusters relative to the number of reconstructed tracks in the full calorimeter acceptance via
  \begin{equation}
    \epsilon_\mathrm{TM} = N_\mathrm{clus}^{\text{matched}} / N_\mathrm{tracks}^{\text{in acc.}},
    \label{eq:tmeffitrk}
  \end{equation}
  or relative to the number of reconstructed clusters via
  \begin{equation}
    \kappa_\mathrm{TM} = N_\mathrm{clus}^{\text{matched}} / N_\mathrm{clus}.
  \end{equation}
  \Figure{fig:tmeffiEBF} shows the track matching efficiencies ($\epsilon_\mathrm{TM}$) for the different calorimeters for single particle simulations of either electrons or charged pions in the full ECCE GEANT4 detector setup.
  For a majority of the ECal acceptance, an excellent efficiency of $\epsilon_\mathrm{TM}>95\%$ is observed.
  Expected deviations towards lower particle momenta are observed, where the track to cluster association breaks down due to a cut-off in the particle cluster reconstruction imposed by the minimum seeding and aggregation thresholds.
  An additional pseudorapdity dependence for the track matching efficiencies is expected due to the previously observed cluster position resolution, which deteriorates for certain $\eta$ regions as wellas the reducted cluster finding efficiency due to remaining acceptance effects. 
  Moreover, the particles might be stronger deflected in the $\eta$-regions with higher amounts of material due to support structures.\\
  Further insights into the track matching efficiency are given by \Fig{fig:tmeffitrkclus}, where the track matching efficiencies $\kappa_\mathrm{TM}$ are shown for all calorimeters in their nominal acceptance.
  The comparison of $\epsilon_\mathrm{TM}$ and $\kappa_\mathrm{TM}$ highlights that the track matching efficiency depends equally on the cluster finding efficiency and the track finding efficiency.
  This can clearly be seen in the electron matching efficiency for the ECals, where $\epsilon_\mathrm{TM}$ is nearly unity when calculated according to \Equation{eq:tmeffitrk}, meaning that if a track is found, it is nearly always matched to a cluster.
  On the other hand, $\kappa_\mathrm{TM}$ shows a reduced efficiency, meaning that for a large portion of clusters no track is found for matching, especially in the forward region.
  For the HCals, the performance is generally worse as particles can pre-shower in the ECals, resulting in clusters with distorted positions on the HCals, thus not for all tracks a matching cluster is found.

\subsection{Particle Identification}
  The information provided by the ECals and HCal can help distinguish between particle species and thus provide highly efficient particle identification, which is crucial for a variety of physics analyses.
  This section therefore focuses first on the PID capabilities of the ECals and subsequently the additional benefits from the HCals.
  The electromagnetic calorimeters (EEMC, BEMC and FEMC) are most commonly used to identify electromagnetic showers coming from a single particle.
  They can differentiate between photons and their background from merged $\pi^0$ decay photons.
  If tracking information is used in addition, the calorimeters can be used to provide a strong separation power between electrons and charged hadrons like $\pi^\pm$, kaons or protons.
  In the following, the different PID approaches are briefly explained and the expected performance is shown based on full detector GEANT4 simulations.
  \begin{figure}[!t] 
    \centering
    \includegraphics[width=0.4\textwidth]{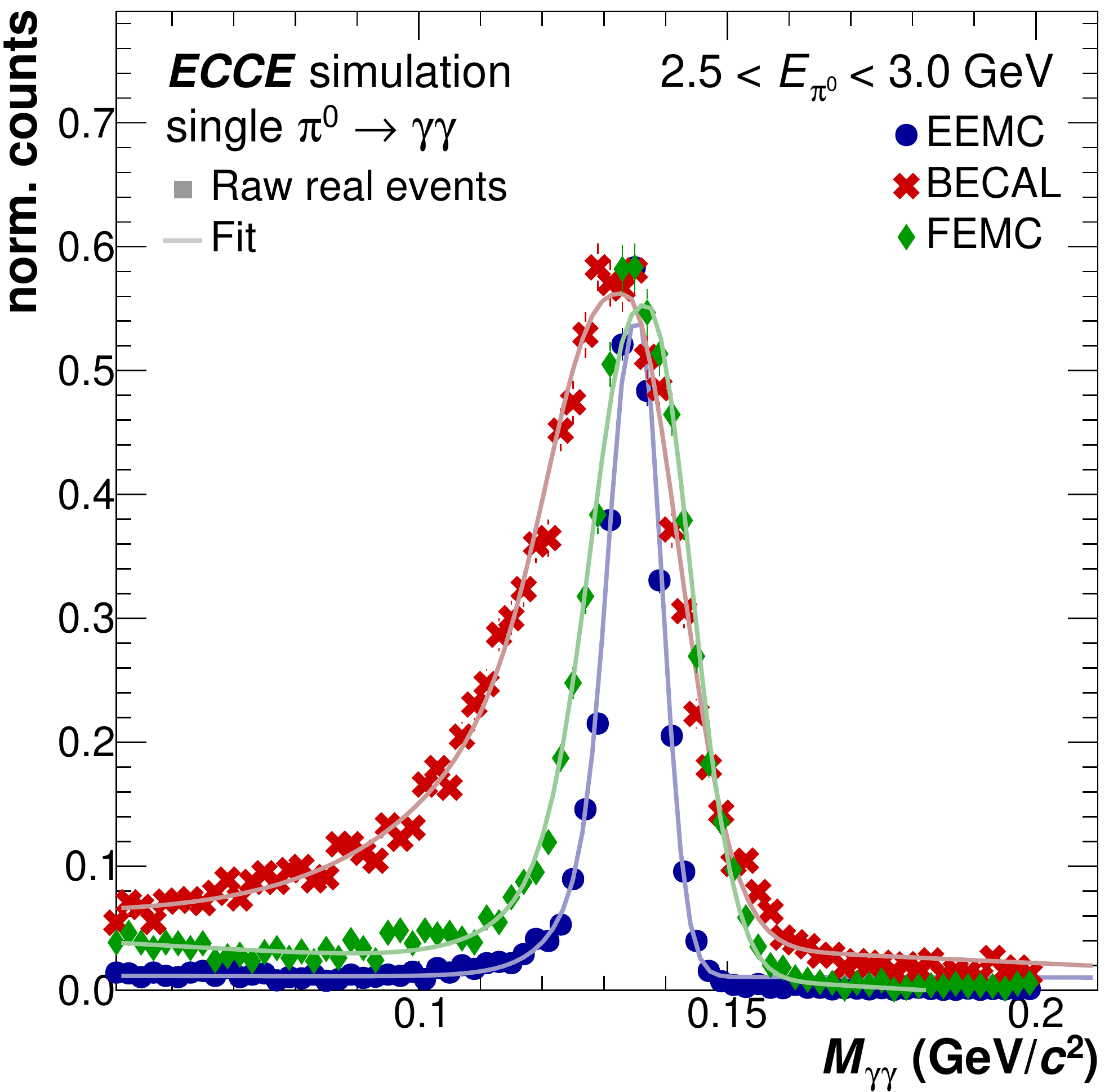} 
    \caption{Invariant mass $M_{\gamma\gamma}$ distribution for generated $\pi^0$ mesons in the energy range from 2.5 to 3.0 GeV for EEMC, BEMC, and FEMC including a composite Gaussian fit function that includes a left-sided exponential tail component.}
    \label{fig:invmasspi0sep}
  \end{figure}
  \begin{figure}[!t] 
    \centering
    \includegraphics[width=0.4\textwidth]{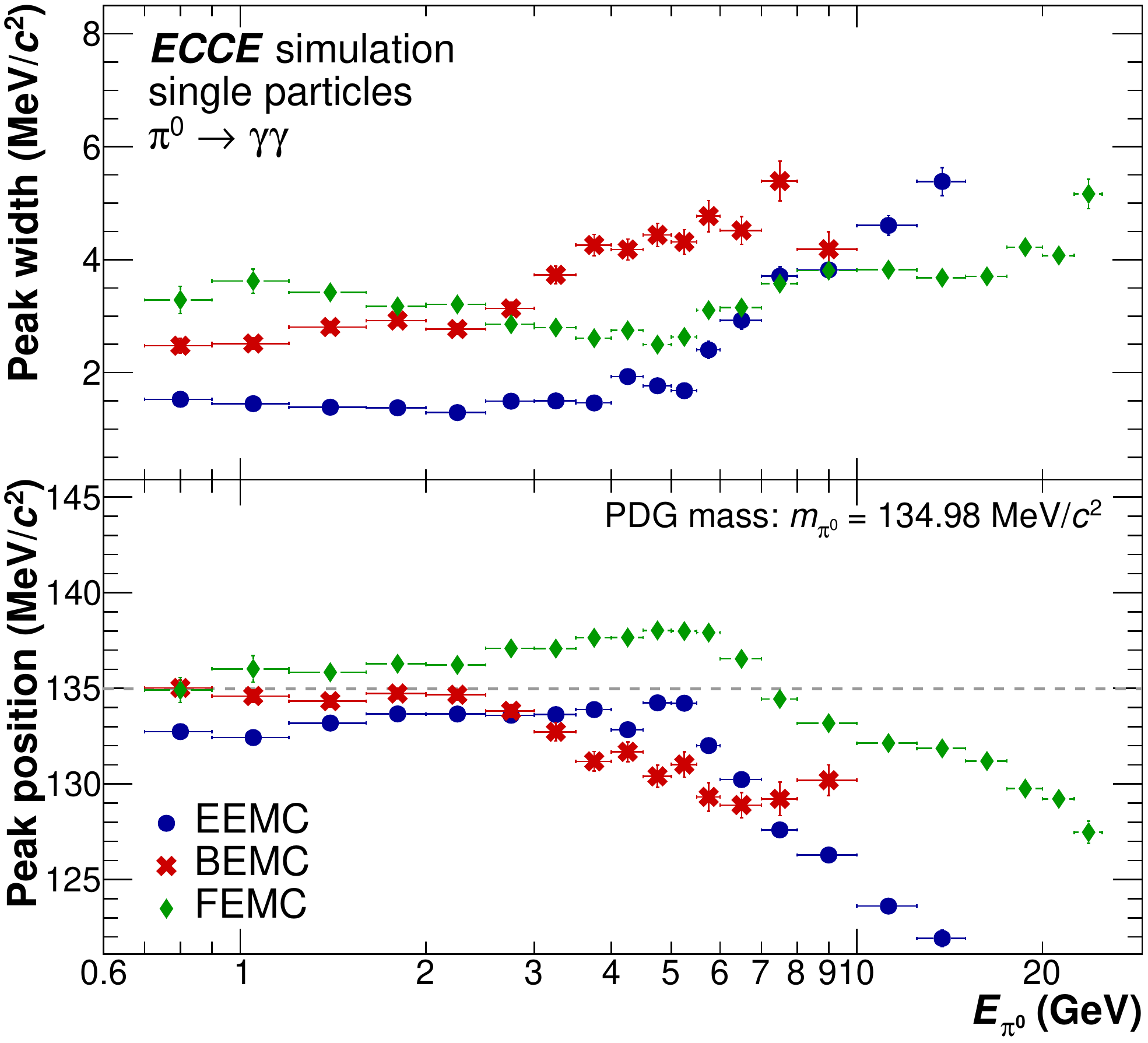}
    \caption{Invariant mass $M_{\gamma\gamma}$ peak width (top) and position (bottom) obtained from a composite Gaussian fit including a left-sided exponential tail.}
    \label{fig:invmassfittedvalues}
  \end{figure}

  \begin{figure*}[!t] 
    \centering
    \includegraphics[width=\textwidth]{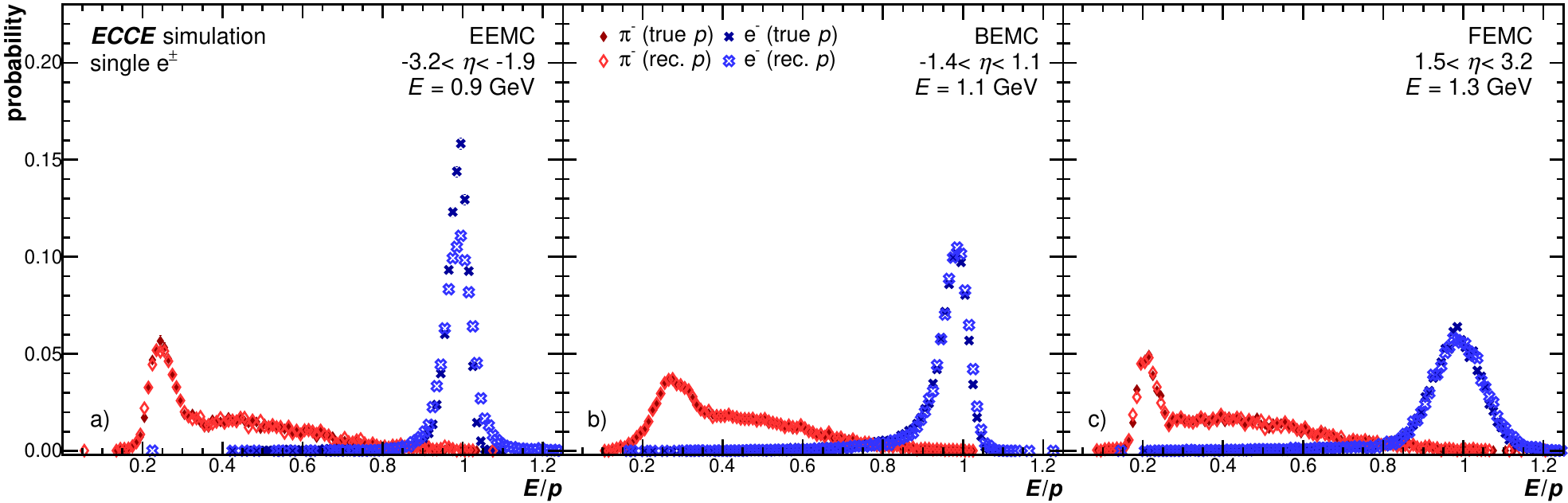}
    \caption{$E/p$ distribution for electrons (blue) and charged pions (red) in the EEMC (right) and the BEMC (left). The $E/p$ distribution is shown for two different approaches where $E/p$ is either calculated using the generated (true) particle momentum or the reconstructed tracking based (rec.) momentum. For both distributions, the full ECCE detector has been simulated using its GEANT4 implementation.}
    \label{fig:eopdistribution}
  \end{figure*}

  \begin{figure}[!ht] 
    \centering
    \includegraphics[width=0.4\textwidth]{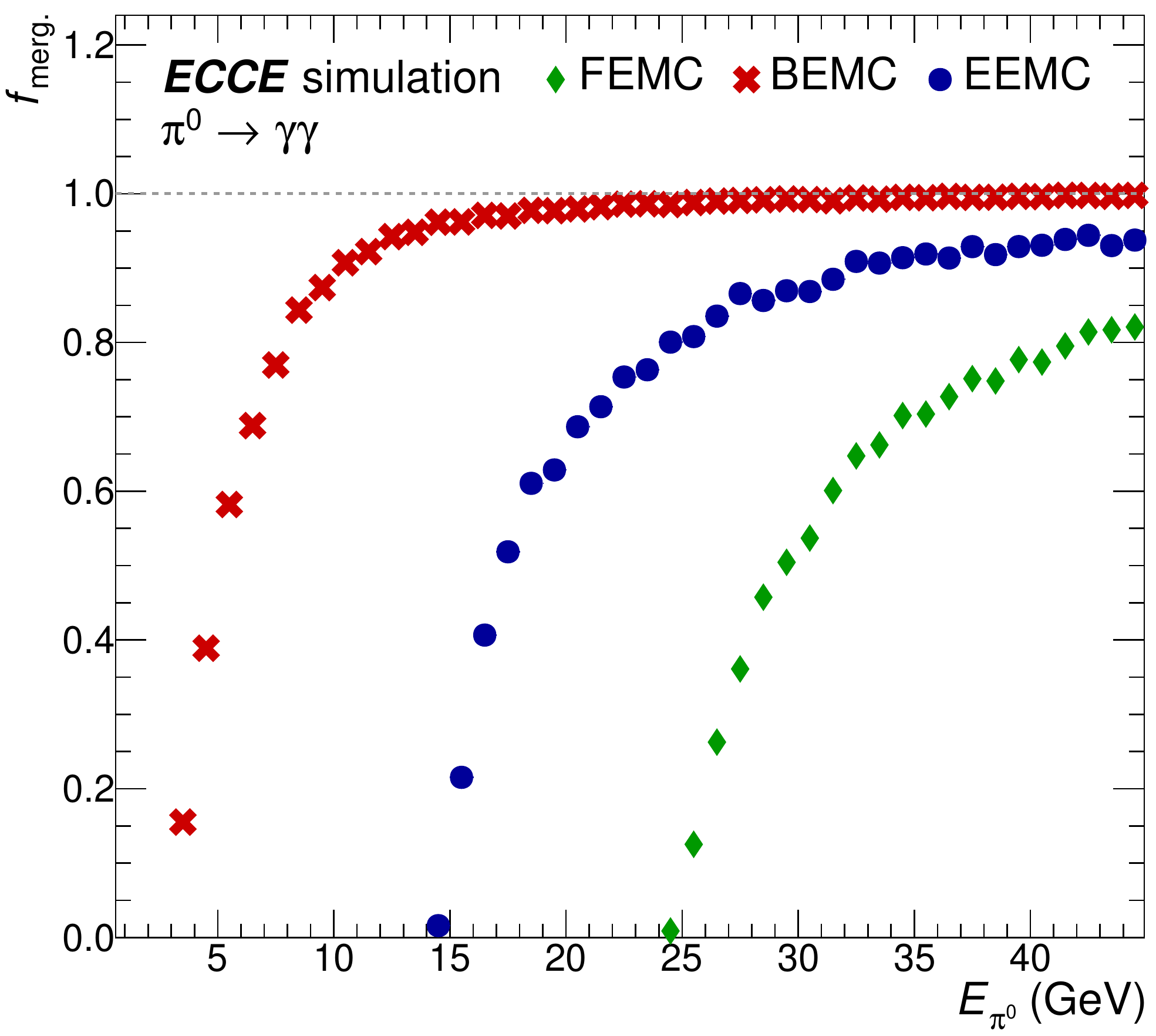}
    \caption{Fraction of neutral pions for which the showers from their decay photons are merged into a single cluster and can not be reconstructed using an invariant-mass-based approach for the different ECals.
    }
    \label{fig:pi0merging}
  \end{figure}

\subsubsection{Single photon and neutral pion separation}
  A significant background for photon analyses originates from $\pi^0$ meson decay photons, which end up in the same reconstructed cluster due to their close proximity.
  In general, when the decay photons can still be reconstructed separately, their calculated invariant mass ($M_{\gamma\gamma}=\sqrt{2E_{\gamma_1}E_{\gamma_2}(1-\cos{\theta_{12}})}$) can be used to veto decay photon clusters if the mass falls in a certain window around the nominal $\pi^0$ mass.
  Example invariant mass distributions for the ECals are shown in \Fig{fig:invmasspi0sep} for a selected energy range of the two-photon meson candidates including a composite Gaussian fit with a left-sided exponential tail.
  The BEMC invariant mass distribution is significantly wider than that of the EEMC or FEMC, as can also be seen in \Fig{fig:invmassfittedvalues}, where the peak width (obtained from the width of the Gaussian fit component) is shown as a function of $\pi^0$ energy.
  The broadening of the peak width with increasing energy and the cutoff of the BEMC data at $E\approx12$ GeV is further elaborated in the following.
  Above a certain energy, the decay kinematics of the $\pi^0$ together with the granularity and resolution of the calorimeter no longer allow to reconstruct separate decay photons due to a merging of their showers.
  Thus, the separation power between meson decay photons and single photons decreases and the reconstructed meson mass starts to deviate from the nominal meson mass as seen in \Fig{fig:invmassfittedvalues}.\\
  The energy dependence of this cluster merging effect is shown in \Fig{fig:pi0merging} for the three ECals.
  The close proximity of the BEMC to the interaction point together with its $4 \times 4$~cm tower size results in a large fraction of merged decay photon clusters already at 5 GeV around $\eta = 0$.
  For $|\eta| > 0.9$ within the BEMC the merging starts to set in at around $10$~GeV due to the larger distance of the calorimeter surface from the interaction point and thus a larger average distance of the two decay photons on the calorimeter surface.
  In contrast, the higher granularity and larger distance from the IP of the EEMC and FEMC, respectively, results in a much later onset of the cluster merging. 
  For the FEMC this effect becomes only significant above 25 GeV, while the EEMC experiences this effect already above 15 GeV.

\subsubsection{Electron PID via charged pion rejection}
Several observables of EIC physics require a clean electron sample \cite{AbdulKhalek:2021gbh}.
One of the largest backgrounds for electrons stems from charged pions ($\pi^\pm$), which can be distinguished on a statistical basis from electrons with a high efficiency using ECal information.
The so-called pion rejection factor is a handle on how strong this $e^\pm$--$\pi^\pm$ separation is for a given calorimeter.
It can be calculated by simulating the response for single electron and separate single pion events.
The quantity $E/p$, meaning the reconstructed cluster energy relative to the incident particle momentum exhibits only slightly overlapping distributions for both particles.
This is shown in \Fig{fig:eopdistribution} where electrons (blue) show a strong enhancement around $E/p\approx1$, while charged pions (red) are smeared towards lower $E/p$ values for all three ECCE ECals.
In realistic events, e.g. based on the Pythia event generator, one expects significantly more hadrons relative to electrons and thus the hadronic tail overlap is expected to be stronger than shown in \Fig{fig:eopdistribution}.
This effect is further enhanced by the presence of jets which result in shower overlaps in the calorimeters.
The track momentum in the following is determined using the full ECCE tracking capabilities \cite{ecce-paper-det-2022-03}.

Due to the small overlap of the $E/p$ distribution for different particle species, a minimum $E/p$ cut can be employed to reject the majority of charged pions in the sample while retaining a high efficiency electron sample.
In previous studies, a minimum cut of $\Delta=1.6\,\sigma_E/E$ has been determined to result in a high electron efficiency of $\varepsilon_e\approx 95\%$. 
However, the asymmetric electron resolution distributions of the calorimeters within the full ECCE integration, as shown in \Fig{fig:eresodistributioneemcbecalfemc} lead to a significantly reduced electron efficiency when applying a 1.6$\sigma$-based cut.
Especially for the BEMC where a strong tail in the energy resolution distribution is visible, the cut results in an electron efficiency of $\varepsilon_e\approx70\%$, while for the other ECals values of about 90--95\% are observed.
Thus, an additional $E/p$ cut value has been determined that allows for $\varepsilon_e= 95\%$, which is indicated as $\varepsilon_{95\%}$ in the following.
This cut value corresponds to approximately $2\sigma$ for the EEMC, $6\sigma$ for the BEMC, and $3\sigma$ for the FEMC, highlighting the difference in the energy resolution peak asymmetry for the various ECCE ECals.
\begin{figure}[!t] 
    \centering
    \includegraphics[width=0.4\textwidth]{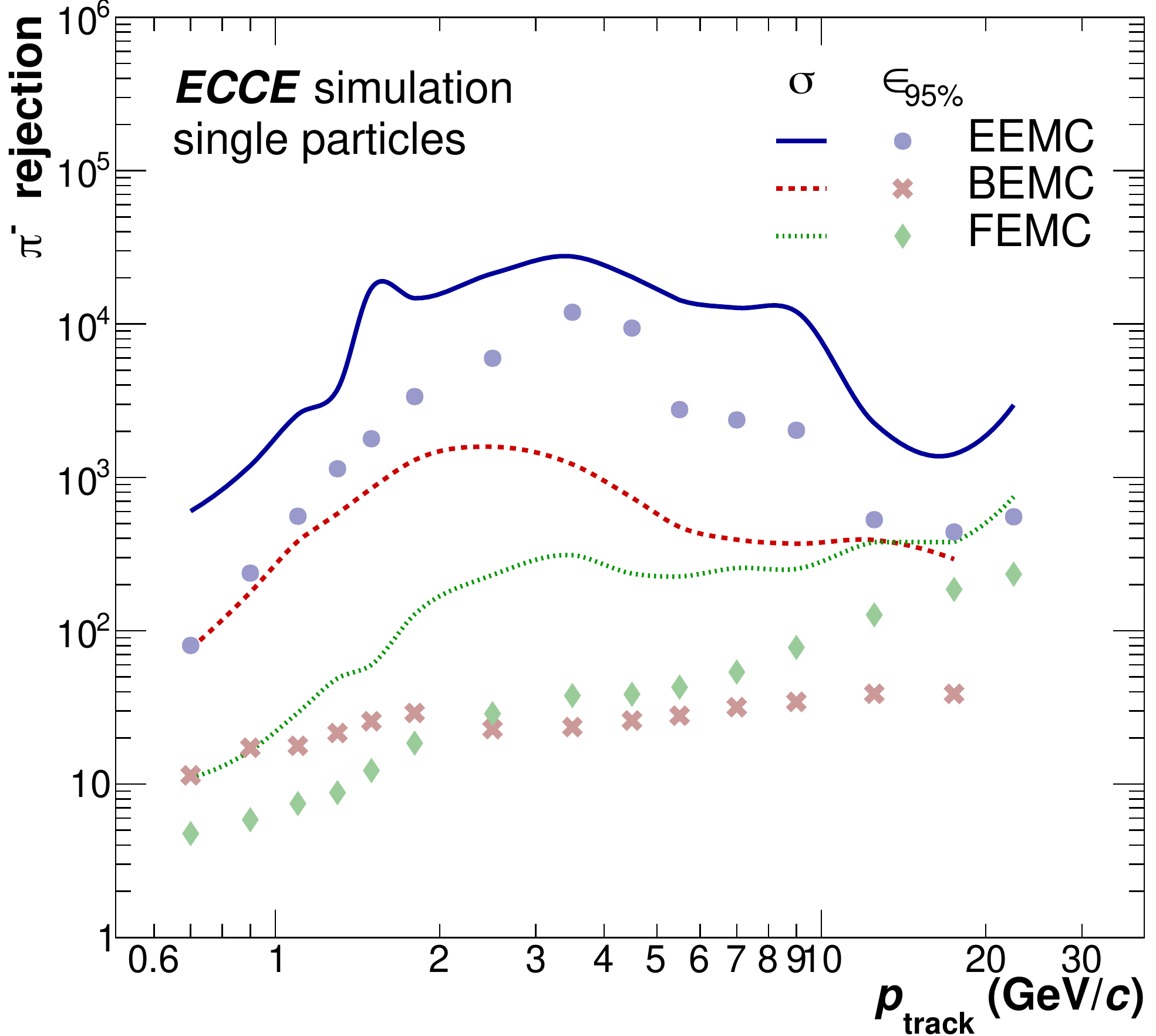}
    \caption{Pion rejection factor for the different ECals with $E/p>1-1.6\,\sigma_e/E$ or based on a $\varepsilon_e\approx 95\%$ cut.
    }
    \label{fig:pionrejection}
\end{figure}
Applying these cuts on the single pion event simulations results in the rejection factors shown in \Fig{fig:pionrejection}.
Values up to $6\times10^4$ are reached for the EEMC using the $1.6\,\sigma$-based cut, while for the other calorimeters $\pi^\pm$ rejection factors ranging from 20 to more than $10^3$ are reached.
For the EEMC the pion rejection capabilities are so striking that an accurate pion rejection factor is hard to determine with the currently available single particle production statistics and the reported values should be interpreted as lower limits.
A significant reduction of about an order of magnitude in the $\pi^\pm$ rejection is observed for the $\varepsilon_e=95\%$ based cut for the FEMC and BEMC, which therefore stands in no reasonable relation to the efficiency loss observed for the other $E/p$ cut values.
This loss mainly arrises from the significant tails observed for these two calorimeters in their current configuration.

\begin{figure}[!t] 
    \centering
    \includegraphics[width=0.4\textwidth]{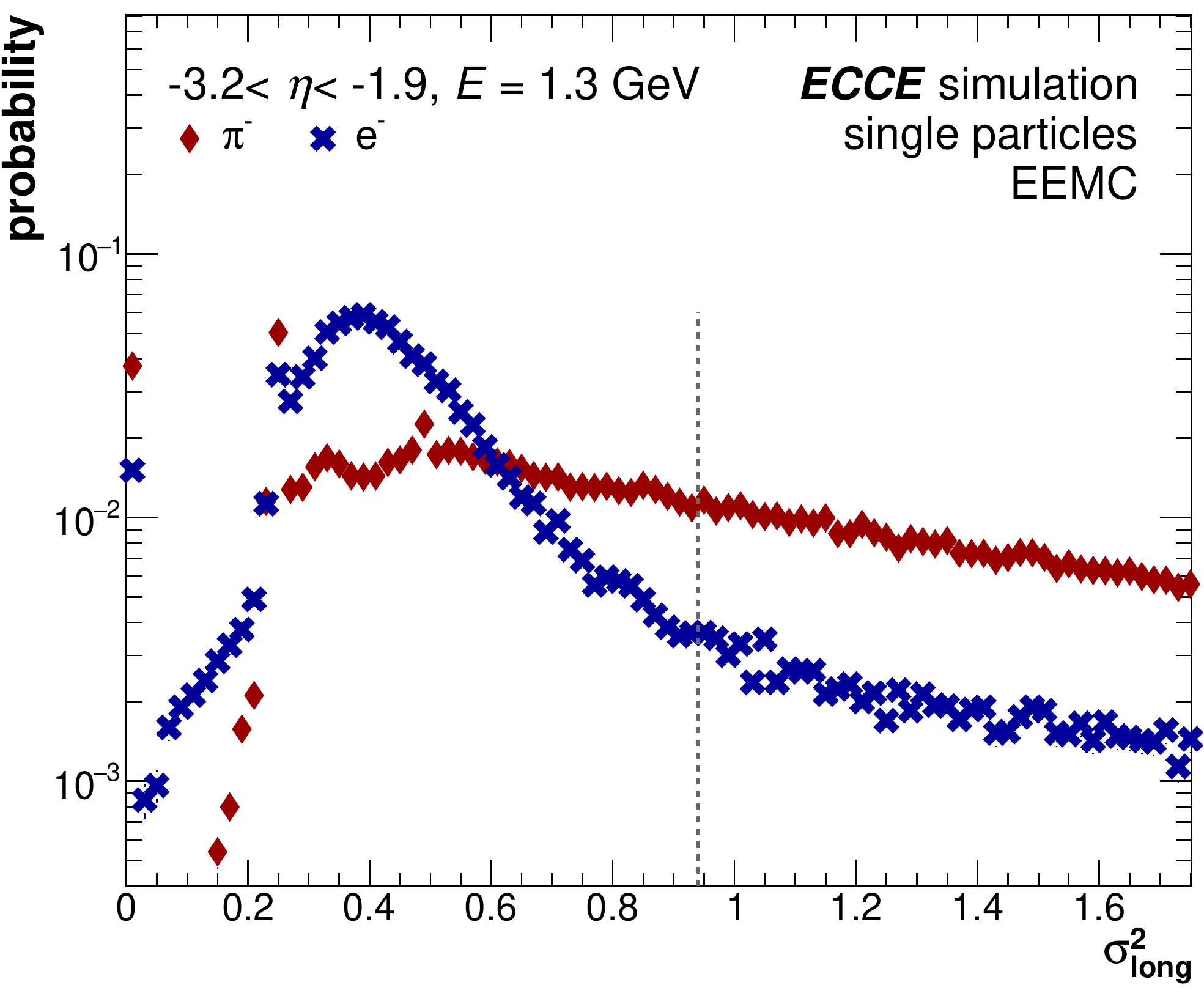}    
    \caption{The shower shape distribution (\shshlo) for electrons (blue) and charged pions (red) in the EEMC using the full ECCE detector simulations at a fixed generated energy of $E=1.3$ GeV. 
             The gray vertical line indicates the \shshlo\ cut value below which 90\% of the electrons would be kept.}
    \label{fig:SSdistribution}
\end{figure}

\subsubsection{Hadron PID}
Besides using an $E/p$ cut to differentiate between electrons and hadrons the shape of the shower and thus the cluster can be used.
The distribution of energy within a cluster is referred to as ``shower shape'', which is described using a parametrization of the shower surface ellipse axes~\cite{Alessandro:2006yt, Abeysekara:2010ze}. 
The shower surface is defined by the intersection of the cone containing the shower with the front plane of the calorimeter. 
The energy distribution along the $\eta$ and $\varphi$ directions is represented by a covariance matrix with terms ${\sigma_{\varphi\varphi}}$, ${\sigma_{\eta\eta}}$ and ${\sigma_{\varphi\eta}}$, which are calculated using logarithmic energy weights $w_i$.
The tower dependent weights are expressed as:
\begin{equation}
\label{eq:w_i}
w_i = {\rm Maximum}(0,w_0+\ln(E_{i}/E_\mathrm{cluster}))
\end{equation}
and
\begin{equation}
w_\mathrm{tot} = \sum_i w_i,\\
\end{equation}
where $w_0$~=~4.5 for the EEMC \cite{Awes:1992yp},  which excludes towers with energy smaller than 1.1\% of the cluster energy.
For the BEMC and FEMC  $w_0$~=~4.0 and $w_0$~=~3.5 are used, respectively, in order to compensate for the different Moliere radii and tower size.
The covariance matrix terms can then be calculated as follows
\begin{equation}
\sigma^{2}_{\alpha\beta} = \sum_i \frac{w_i\alpha_i\beta_i}{w_\mathrm{tot}}-\sum_i \frac{w_i\alpha_i}{w_\mathrm{tot}}\sum_i \frac{w_i\beta_i}{w_\mathrm{tot}}\,,
\label{eq:ss_centroid}
\end{equation}
where $\alpha_{i}$ and $\beta_{i}$ are the tower indices in the  $\eta$ or $\varphi$ direction. 
Similarly, also the average cluster position in the $\eta$ and $\varphi$ direction in the calorimeter plane is determined using the tower positions weighted logarithmically by their deposited energy \cite{Awes:1992yp}.\\  
The shower shape parameters \shshlo\ (long axis) and \shshsh\ (short axis) are defined as the eigenvalues of the covariance matrix, and are calculated as
\begin{eqnarray}
\shshlo = 0.5(\sigma^{2}_{\varphi\varphi}+\sigma^{2}_{\eta\eta})+\sqrt{0.25(\sigma^{2}_{\varphi\varphi}-\sigma^{2}_{\eta\eta})^2+\sigma^{2}_{\eta\varphi}}, \label{eq:ss1}\\
\shshsh = 0.5(\sigma^{2}_{\varphi\varphi}+\sigma^{2}_{\eta\eta})-\sqrt{0.25(\sigma^{2}_{\varphi\varphi}-\sigma^{2}_{\eta\eta})^2+\sigma^{2}_{\eta\varphi}} \label{eq:ss2},
\end{eqnarray}
Previous experiments have determined that the short axis \shshsh\ carries significantly less discriminative power compared to \shshlo\ and thus only the long axis is considered in the following.

\begin{figure}[!htb] 
    \centering
    \includegraphics[width=0.4\textwidth]{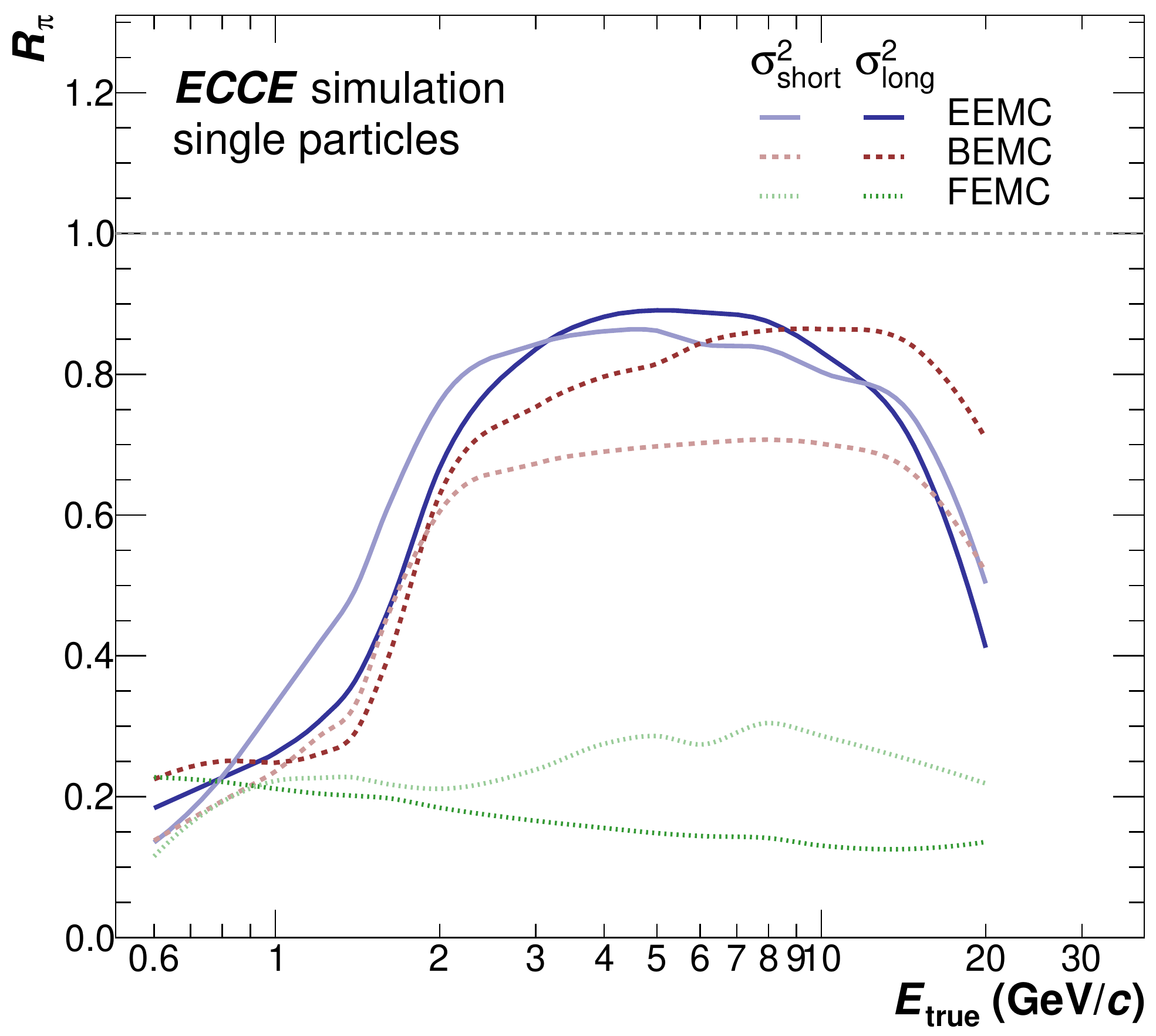}
    \includegraphics[width=0.4\textwidth]{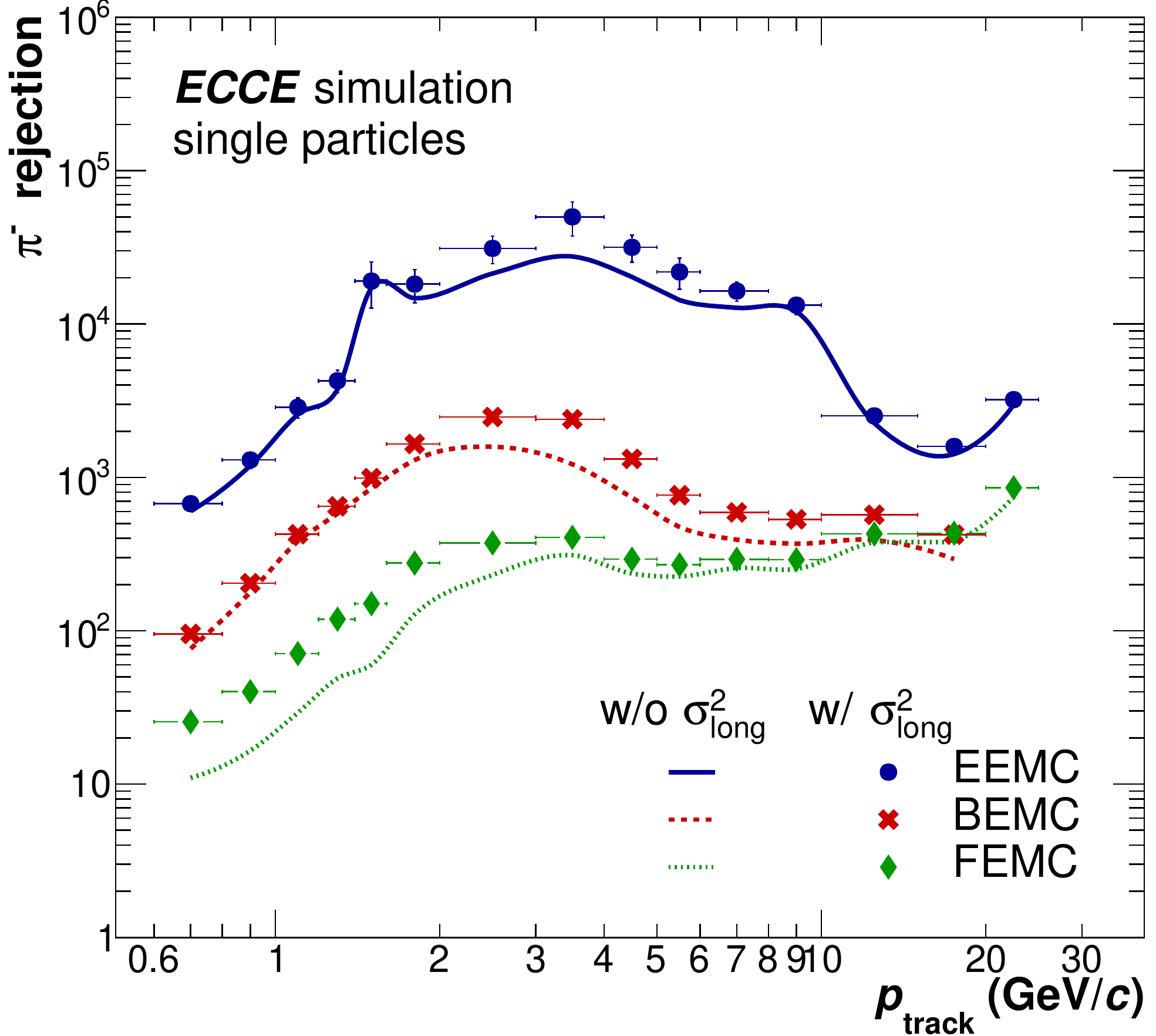}
    \caption{Left: Fraction ($R_\pi$) of cluster originating from charged pions, which can be rejected due to the chosen shower shape cut as a function of the incident pion energy. 
             Right: Pion rejection factor for the different ECals with $E/p>1-1.6\,\sigma_e/E$ (w$/$o PID) or $E/p>1-1.6\,\sigma_e/E$ and the additional \shshlo\ selection (w$/$ PID) applied as a function of the true track momentum.}
    \label{fig:RejectionPID}
\end{figure}
Using these parameters symmetric electromagnetic showers with a small spread originating either from photons or electrons can be distinguish from non-symmetric showers caused by hadronic interactions.
The shower shape of charged particles can also be elongated by the angle of incidence.  calorimeters.
Furthermore, the merging of showers from electromagnetic processes, i.e. ${\rm e^{+}e^{-}}$ pairs from conversions within a close distance to the calorimeter or photons from neutral meson decays with high transverse momenta, also lead to asymmetric shower shapes.\\
An example distribution of the shower shape parameter \shshlo\ for electrons (blue) and pions (red) as seen by the EEMC can be found in \Fig{fig:SSdistribution}.
As can be seen, the energy deposits from an electrons at the same incident energy are significantly more collimated than those of charged pions.
Consequently, electron clusters have predominantly lower shower shape values. 
As these distributions strongly change as a function of the incident energy a \shshlo\ cut value function is calculated that preserves 90\% of the electrons. 
Using these cut values based on the shower shape alone up to 90\% of the pions can be rejected in the EEMC as shown in  \Fig{fig:RejectionPID}~(top).
By simultaneously using the aforementioned $E/p$ and \shshlo\ cuts, the pion rejection quoted in \Fig{fig:pionrejection} is improved by at least a factor two in most momentum bins as seen in \Fig{fig:RejectionPID}~(bottom).

\section{Summary}
\label{summary}

In summary, the ECCE calorimeter systems have been designed to support the full scope of the EIC physics program as presented in the EIC white paper~\cite{Accardi:2012qut} and in the 2018 report by the National Academies of Science (NAS)~\cite{NAP25171}. These systems can be built within the budget envelope set out by the EIC project while simultaneously managing cost and schedule risks.

\section{Acknowledgements}
\label{acknowledgements}

We acknowledge support from the Office of Nuclear Physics in the Office of Science in the Department of Energy, the National Science Foundation, and the Los Alamos National Laboratory Laboratory Directed Research and Development (LDRD) 20200022DR.

We thank (list of individuals who are not coauthors) for their useful discussions and comments.


\bibliographystyle{elsarticle-num} 
\bibliography{refs.bib,refs-ecce.bib}

\end{document}